\documentclass[twocolumn]{aastex62}
\usepackage[T1]{fontenc}
\usepackage{color,natbib,graphicx,amsmath,amsfonts}
\newcommand{\epk}{2004.532}
\newcommand{\LPR}{8e:4d:2c:1b}
\renewcommand{\arraystretch}{1.2}
\def\idm#1{{\mbox{\scriptsize #1}}}
\def\vec#1{{\pmb #1}}

\newcommand{\au}{\mbox{au}}
\newcommand{\msun}{\mbox{M}_{\odot}}

\newcommand{\mJ}{\,\mbox{m}_{\idm{Jup}}}
\newcommand{\Mmean}{\mathcal{M}}
\newcommand{\Y}{\langle Y \rangle}
\newcommand{\Ym}{\langle Y \rangle}
\newcommand{\mmegno}{\left|\langle Y \rangle -2 \right|}
\newcommand{\Chi}{\chi^2_{\nu}}
\def\moa{{MCOA}}
\def\megno{{MEGNO}}
\def\hr8799{{HR~8799}}
\def\mercury{{\tt Mercury 6.3}}
\def\emcee{{\tt emcee}}
\def\mufarm{{\tt $\mu$Farm}}

\begin{document}

\shorttitle{The orbital architecture and debris disks in the HR 8799 planetary system}
\shortauthors{Go\'zdziewski \& Migaszewski}
\title{The orbital architecture and debris disks of the HR 8799 planetary system}
\author[0000-0002-0786-7307]{Krzysztof Go\'zdziewski}
\author{Cezary Migaszewski}
\affil{
Faculty of Physics, Astronomy and Informatics, Nicolaus Copernicus University, Grudzi\c{a}dzka 5, 87-100 Toru\'n, Poland
}
%
\begin{abstract}
The \hr8799{} planetary system with four $\simeq 10 \mJ$ planets in wide orbits up to $\simeq 70$~au, and orbital periods up to 500~yrs has been detected with the direct imaging. Its intriguing orbital architecture is not yet fully resolved due to time-limited astrometry covering only $\simeq 20$~years. Earlier, we constructed a heuristic model of the system based on rapid, convergent migration of the planets. Here we developed a better structured and CPU-efficient variant of this model. With the updated approach, we re-analyzed the self-consistent, homogeneous astrometric dataset in \citep{Konopacky2016}. The best-fitting configuration agrees with our earlier findings. The \hr8799{} planets are likely involved in dynamically robust Laplace \LPR{} resonance chain. Hypothetical planets with masses below the current detection limit of 0.1-3~Jupiter masses, within the observed inner, or beyond the outer orbit, respectively, do not influence the long term stability of the system. We predict positions of such non-detected objects. The long-term stable orbital model of the observed planets helps to simulate the dynamical structure of debris disks in the system. A CPU-efficient fast indicator technique makes it possible to  reveal their complex, resonant shape in $10^6$ particles scale. We examine the inner edge of the outer disk detected between $90$ and $145$~au. We also reconstruct the outer disk assuming that it has been influenced by convergent migration of the planets. A complex shape of the disk strongly depends on various dynamical factors, like orbits and masses of non-detected planets. It may be highly non-circular and its models are yet non-unique, regarding both  observational constraints, as well as its origin.
\end{abstract}
%
\section{Introduction}
\label{sec:introduction}
%

The \hr8799{} planetary system has been discovered by \cite{Marois2008} as a three-planet configuration. Shortly, after two years, the fourth innermost planet has been announced by the same team \citep{Marois2010}. Since the discovery, this unusual extrasolar planetary system was studied in literally tens of papers. The parent star age, companions masses, as well as the orbital architecture and its long-term stability, debris disks, and formation are analyzed in, e.g., \cite[][]{Read2018,Wilner2018,Wertz2017,Gotberg2016,Booth2016,Konopacky2016,Contro2016,Currie2016,Maire2015,Zurlo2016,Pueyo2015,Matthews2014,Marleau2014,
Gozdziewski2014A,Oppenheimer2013,Baines2012,Esposito2013,Currie2012,Sudol2012,Soummer2011,Bergfors2011,Currie2011,Galicher2011,Hintz2010,Marshall2010,MoroMartin2010,Metchev2010,Fabrycky2010,Lafreniere2009,Gozdziewski2009,Su2009,Reidemeister2009},
and counting. 

Regarding a characterization of the parent star and its planets, as well as the past and to date observations of debris disks in the system, we refer to the recent papers by \cite{Booth2016}, \cite{Read2018}, \cite{Wilner2018}, and references therein. Following these works, the star is very young of $\sim 60$~Myr, within the most likely interval of 30~Myr to 160~Myr. We adopt its  mass $m_{\star}$ of $1.52\,\msun$, and masses of the planets in the range of $5$--$10\,\mJ$ \citep{Marois2010}.

{
In spite of the enormous literature on the \hr8799{} system, many questions are still open. The global, orbital structure of the \hr8799{} system is a particularly interesting problem.} The ratio of $\simeq 120$ $(\alpha,\delta)$ measurements in the literature, collected in \citep{Wertz2017}, to 24 geometrical elements (free parameters) is 4--5 and it did not significantly change for a few past years, since the discovery of the forth companion in \citep{Marois2010}. Moreover, the major limitation of the astrometric models is a small coverage of the orbits by the measurements, between $\simeq 3\%$ and $\simeq 12\%$ for planets \hr8799{}b and \hr8799{}e, respectively, given significant uncertainties of $\simeq 10$~mas. These weak observational constraints permit a variety of non-unique orbital geometries, although all of them seem equally good fit the observations  \citep[e.g.][]{Wertz2017,Konopacky2016}. 

A dynamical analysis of the best-fitting, Keplerian solutions reveal that they represent crossing orbits and configurations unstable in an enormously short 0.5--1~Myr time scale.  Simple and natural requirements of the stability, like the Hill criterion, do not constrain the orbital models either, e.g., \citep[][their Fig. 3]{Konopacky2016} and \citep[][their Fig. B.1, especially panel for planet~d]{Wertz2017}. Finding {\em long-term stable} configurations simultaneously fulfilling astrometric and mass constraints is difficult even for the lower limit of the star age of 30~Myr and the low limit of the planet masses. 
{
It is still uncertain whether or not the system is strongly resonant, long-term  or only marginally stable, or unstable at all \citep[e.g.,][]{Gotberg2016, Gozdziewski2014A,Gozdziewski2009,Fabrycky2010}. It may remain a matter of somehow philosophical debate unless the orbits are observationally sampled for a sufficiently long interval of time.} 
Furthermore, it is unclear how the system has been formed \citep{Marois2010}, given that massive planetary or brown-dwarf companions are found relatively close to the star. 

Regarding dynamical arguments, none of analytic or semi-analytic criteria of stability apply to the \hr8799{} system. The early dynamical studies of the three-planet system \citep{Fabrycky2010,Gozdziewski2009,Reidemeister2009,Marois2008} revealed that even quasi-circular and apparently wide $\simeq 70$~au orbits are separated by less than 3--4~mutual Hill radii. Such configurations are predicted as self-destructing statistically in a fraction of 1~Myr time-scale \citep{Morrison2016,Chambers1996,Chatterjee2008} unless a protecting dynamical mechanism is present. Indeed, coplanar, or close to coplanar orbits of three outer planets involved in a stable Laplace 4d:2c:1b~MMR were found, explaining the astrometric observations of the \hr8799 system, shortly after its discovery \citep{Fabrycky2010,Gozdziewski2009,Reidemeister2009,Soummer2011,Marshall2010}.

However, the stability problem became much harder with the fourth planet announced in \citep{Marois2010}. Even assuming a protecting MMR mechanism, the orbital parameters of long-living configurations may vary only within small limits \citep{Gozdziewski2014A} or must be particularly tuned, since they are extremely chaotic and prone to tiny changes of the initial conditions and the numerical integrator scheme \citep{Gotberg2016}.

{Seeking for long-term stable orbits of the planets has a ``practical'' aspect, since they are needed for simulating debris disks in the system \citep{Contro2016,Booth2016,Read2018,Wilner2018}}. The outer debris disk might be present between $\simeq 90$~au and $\simeq 450$~au. According to models of the ALMA observations in \citep{Booth2016} and \citep{Read2018}, the inner edge of this disk should be placed at $\sim 145$~au, beyond the direct influence attributed to planet \hr8799{}b roughly at $\sim 90$~au. It might indicate the presence of an additional fifth planet, below the current detection limit of a few Jupiter masses. However \cite{Wilner2018} argue that combined ALMA and VLA observations with higher spatial resolution do not favour the fifth planet hypothesis, and the inner border of the disk may be detected at $\simeq 104$~au, consistent with the  currently known four planet configuration. Moreover, they constrained the outermost planet mass  $m_{\rm b}\simeq 6^{+7}_{-3}\mJ$.
 
We note that some of the works devoted to debris disks \citep{Booth2016,Read2018,Contro2016} made use of our best-fitting model representing the Laplace MMR chain \citep{Gozdziewski2014A} which was found with observations up to the epoch of 2013. The most recent papers regarding astrometric models of the \hr8799{} system based on to date observations, till epoch 2014.93, focus mostly on Keplerian solutions \citep{Wertz2017,Konopacky2016}.  These best-fitting, or most likely models, in terms of the Bayesian inference, exhibit different geometries, such as non-coplanar, highly eccentric non-resonant or partly resonant orbits but have not been examined for their long-term orbital stability.  Therefore such solutions are not suitable for the dynamical analysis of the debris disks. 

{In this work, we} propose to resolve the structure of these disks with CPU efficient fast indicators, based  on the maximal Lyapunov Characteristic Exponent (mLCE), instead of integrating orbits for a required full time span. For that purpose we need to construct long-term, rigorously stable {orbits of the planets} which are robust against small perturbations. Such stable planetary models are helpful to localize ``missing'' (non-detected) planets, or to investigate influence of such putative planets on the debris disks.

This paper is structured as follows. After this Introduction, Section~\ref{sec:mcoa} presents an update of the model of astrometric data through constraining it by the process of planetary migration. Section~\ref{sec:best} regards the results and details of orbital architectures of the HR~8799 system derived for a self-consistent set of astrometric measurements in \citep{Konopacky2016} made with the Keck~II telescope. The results in this part support our idea of the orbital analysis, or could be at least an alternative and reasonable approach when compared with other solutions in the literature.  Section~\ref{sec:calibration} is for our model of debris disks based on the fast indicator technique, and we describe the results of its time-calibration with the direct numerical integrations. Simulations of yet  undetected planets in the system are described in~Sects.~\ref{sec:inner} and~\ref{sec:planetV}.
We analyse the dynamical structure of the inner and outer debris disks in Sect.~\ref{sec:inner} and~\ref{sec:outer}, respectively.  We also investigate a scenario of the outer disk influenced by migrating planets in Sect.~\ref{sec:migration}. Some apparent discrepancies between our approach and the results in the recent literature are addressed in Sect.~\ref{sec:discussion}. We summarize the work in Section~\ref{sec:conclusions}.
%
%
\section{The migration algorithm revisited}
%
\label{sec:mcoa}
The phase space of mutually interacting planetary systems has a discrete, non-continuous structure \citep{Malhotra1998}. This feature may be useful to introduce implicit constraints on the otherwise huge multidimensional space of free parameters. The key idea relies in the evolution of the orbital elements in  migrating planetary systems. Due to the planetary migration, these elements may be tightly self-constrained, depending on an established mean motion resonance (MMR). We assume that such, although not necessarily realistic nor fully resolved dynamical process, orders the planetary system, and drives it to an equilibrium state. In such a state the orbital elements, like the semi-major axes, orbital phases and eccentricities are limited to certain, narrow ranges. This coherence is crucial since it also provides the long-term dynamical stability.

The essential optimization problem is to find MMR-trapped systems which reproduce the observations at some time. We solved it with the Migration Constrained Optimization Algorithm (MCOA), as dubbed earlier in \citep{Gozdziewski2014A}. The \moa{} makes use of a heuristic model of the planetary migration \citep{Beauge2006,Moore2013} and theoretical estimates of the planetary  masses,  consistent with the recent cooling theory \citep{Marois2010,Baines2012,Marleau2014}. 

\begin{figure}
\centerline{
\hbox{\includegraphics[width=0.48\textwidth]{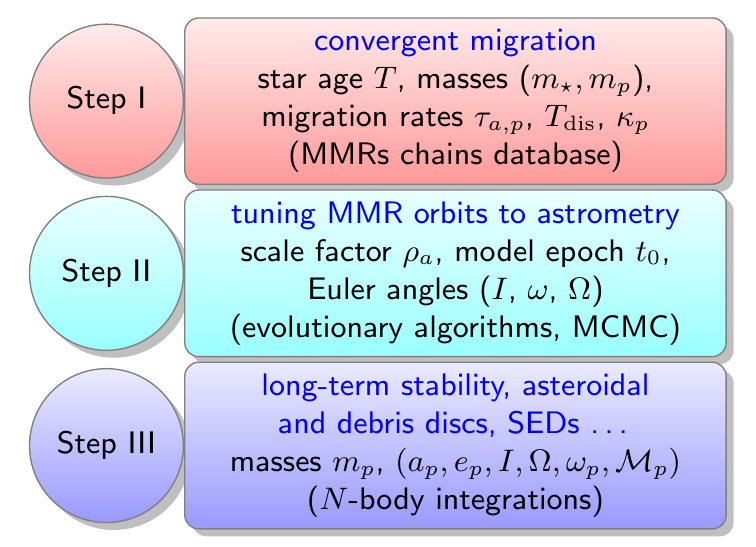}}
}
\caption{
A graphical representation of the Migration Constrained Optimization Algorithm. Major steps and the most important parameters are listed. The diagram illustrates also a plan of this work. See the text for details.
}
\label{fig:fig1}
\end{figure}

The original MCOA would require repeating CPU demanding computations if the astrometric data significantly change. A recent publication of revised and unified astrometric measurements in \citep{Konopacky2016} inspired us to search for CPU efficient, and perhaps improved implementation of the method.  Indeed, we developed a better structured computational strategy which consists of two essentially independent steps. These steps may be conducted standalone, instead of running the original monolithic code. The updated scheme, much easier to follow and repeat, if necessary, is illustrated in Fig.~\ref{fig:fig1} and described below. (Fig.~\ref{fig:fig1} may be considered as a graphical plan of this paper).
\subsection{A set of MMR captured systems}
At the first step, we build a database of systems formed through migration of an appropriate number of planets. Their semi-major axes may be only roughly consistent with the observations. We consider co-planar systems, following arguments behind the planetary migration theory \citep[e.g.,][and references therein]{Armitage2018}. Orbital elements such as eccentricity, nodal angle and the mean anomaly are self-consistently tuned by the migration. The planetary masses may be constrained by cooling models \citep{Baraffe2003,Marleau2014} or sampled from a preselected distribution. 

We may consider any reasonable variant of the migration theory at this stage. The crucial point is that the planetary migration leads to the MMR capture, and establishes stable systems. In order to mimic the migration with the $N$-body code, we modify the astrocentric equations of motion with a force term \citep{Papaloizou2006,Beauge2006,Moore2013} 

\begin{equation}
\vec{f}_i = -\frac{\vec{v}_i}{2\tau_i} - \frac{\vec{v}_i - \vec{v}_{c,i}}{\kappa_i^{-1} \tau_i},
\label{eq:migration}
\end{equation}
where $\vec{v}_i$ is the astrocentric velocity of planet~$i=1,2,3,\ldots,N$, $\vec{v}_{c,i}$ is a velocity of planet~$i$ at a circular Keplerian orbit at the distance of this planet. We note that the \hr8799{} planets are numbered with $i=1,2,3,\ldots$ with respect to their increasing distance from the star, or, we mark them with Roman letters, from the innermost ``e'' $\equiv$ ``1'' to the outermost ``4'' $\equiv$ ``b'' or ``f'' $\equiv$ ``5'', following the order in which they were discovered and named \citep{Marois2008,Marois2010}.

The timescale of migration of planet~$i$ is denoted with $\tau_i$, while $\kappa_i$ is the ratio between $\tau_i$ and the timescale of orbital circularization, which {may} be uniform for all planets in the system. Moreover, $\tau_i$ {could} depend on time, that simulates a dispersal of the disc, i.e., $\tau_i = \tau_i(t=0) \exp(t/T_{\rm dis})$, where $T_{\rm dis}$ is the characteristic time of the decay.

\begin{figure*}
\centerline{ 
\hbox{
\vbox{
\hbox{\includegraphics[height=0.16\textheight]{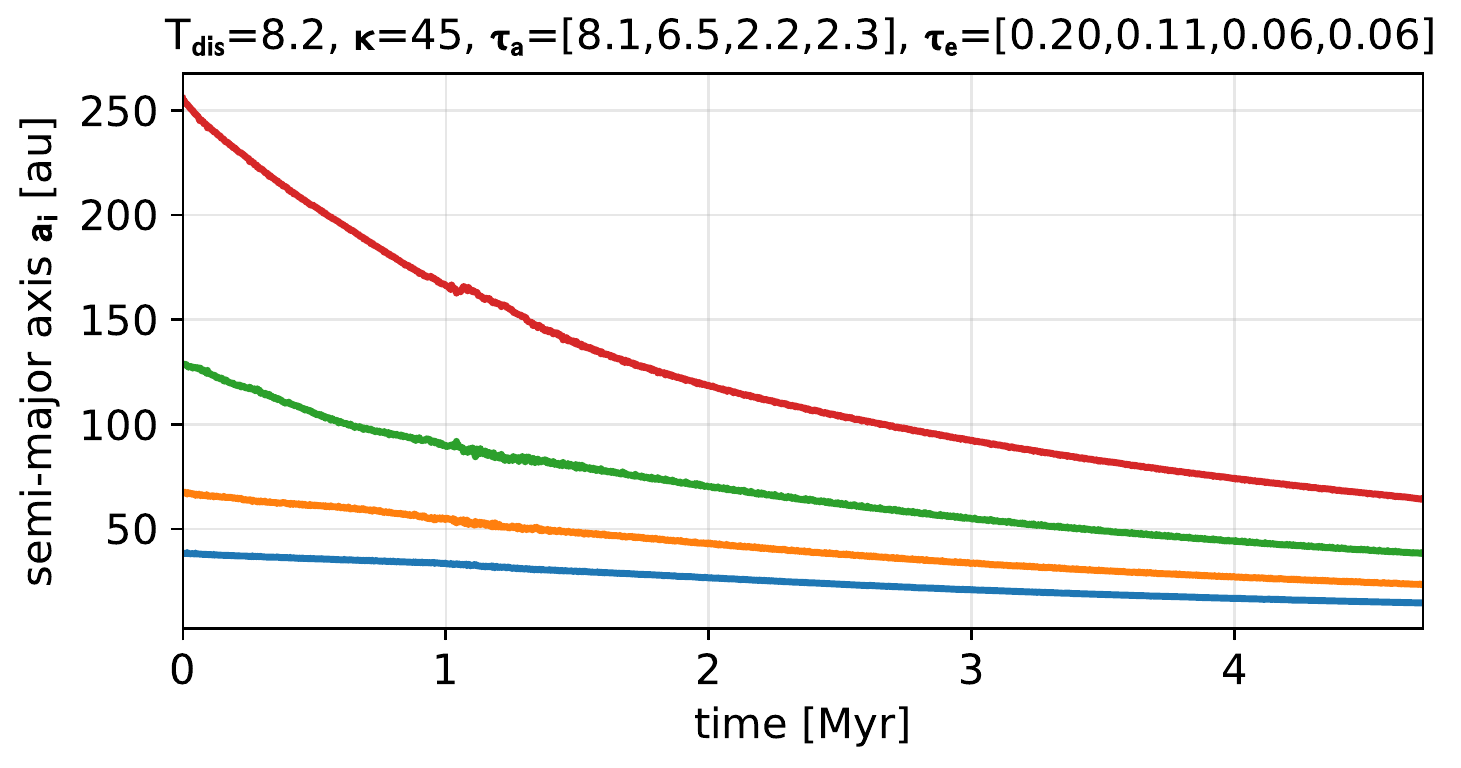}}
\hbox{\includegraphics[height=0.16\textheight]{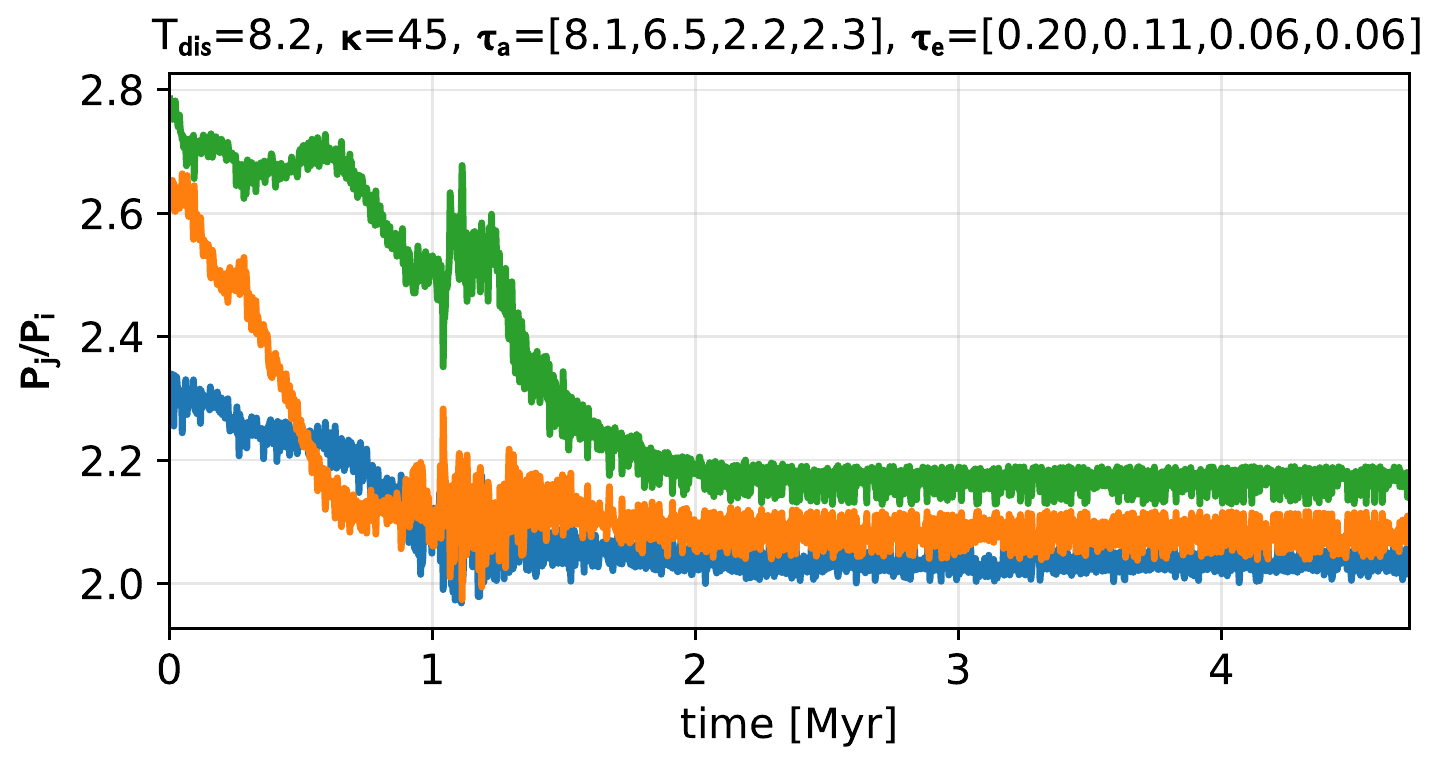}}
\hbox{\includegraphics[height=0.16\textheight]{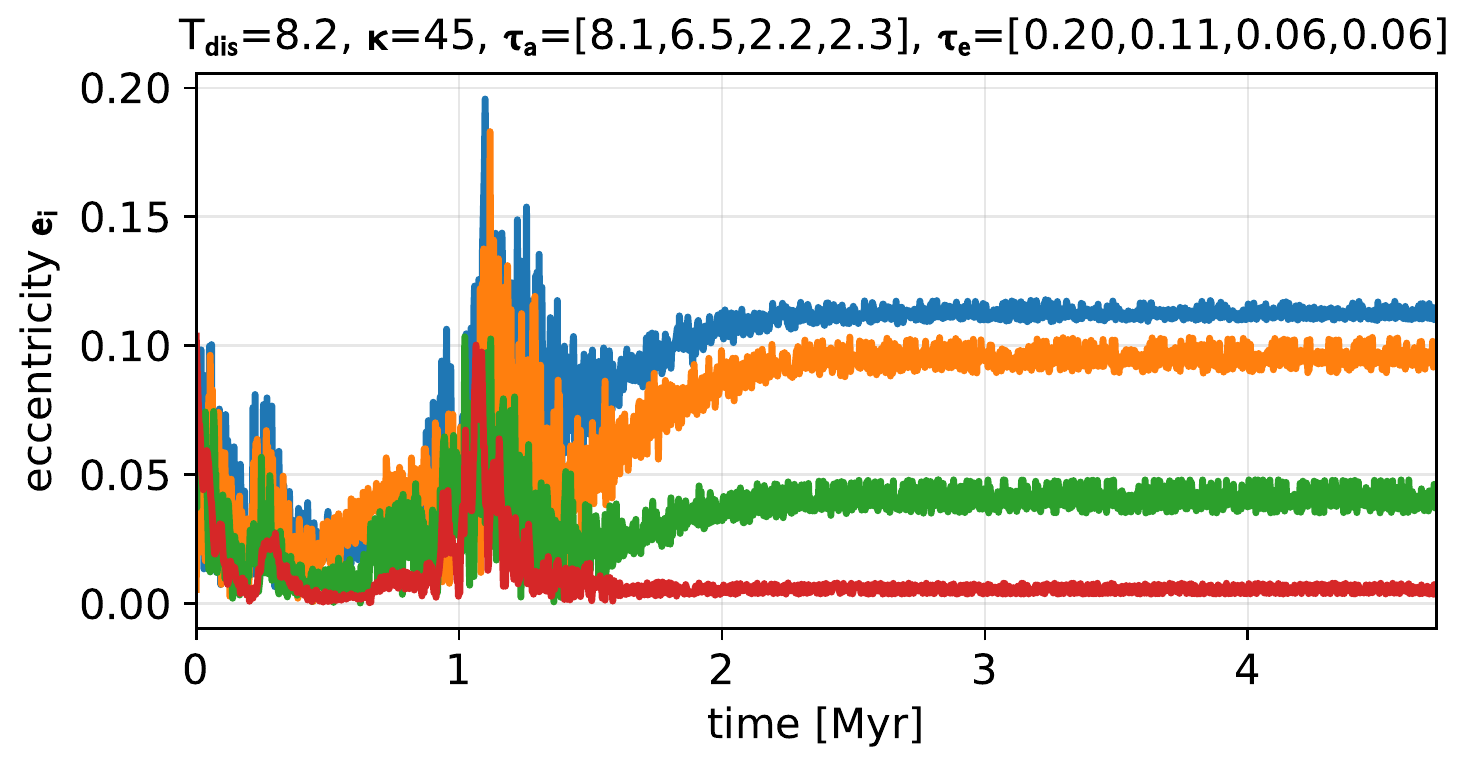}}
\hbox{\includegraphics[height=0.16\textheight]{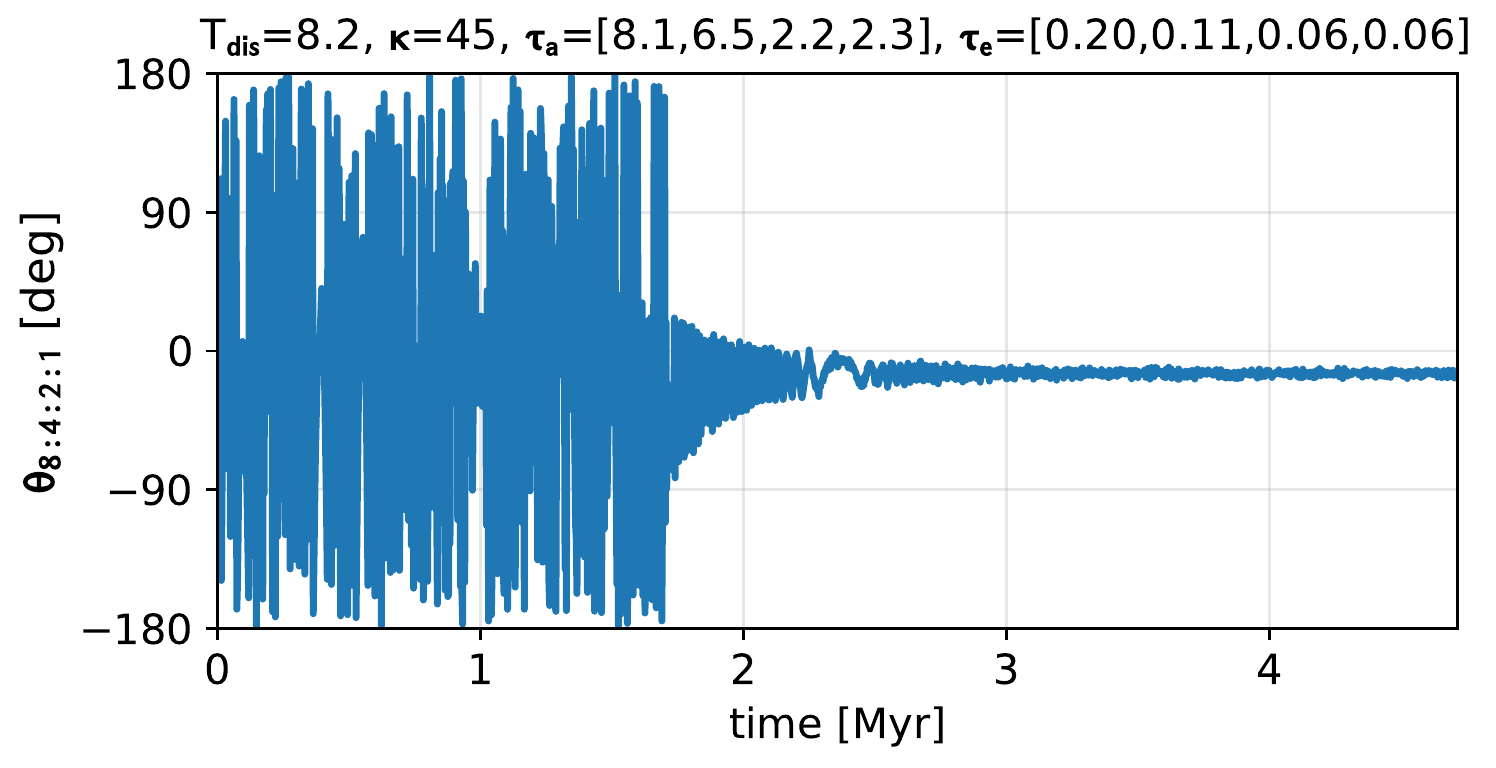}}
\hbox{\includegraphics[height=0.16\textheight]{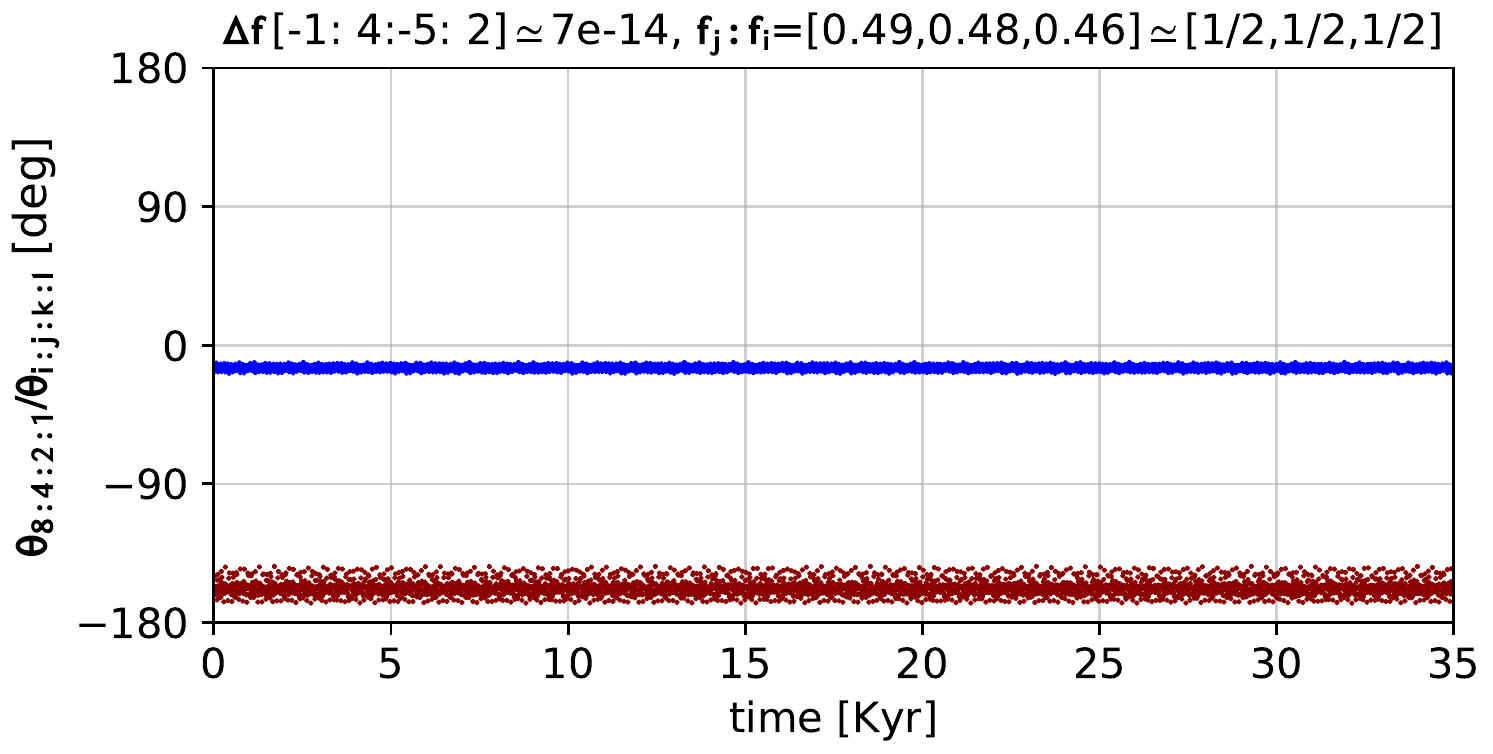}}
}
\qquad
\vbox{
\hbox{\includegraphics[height=0.16\textheight]{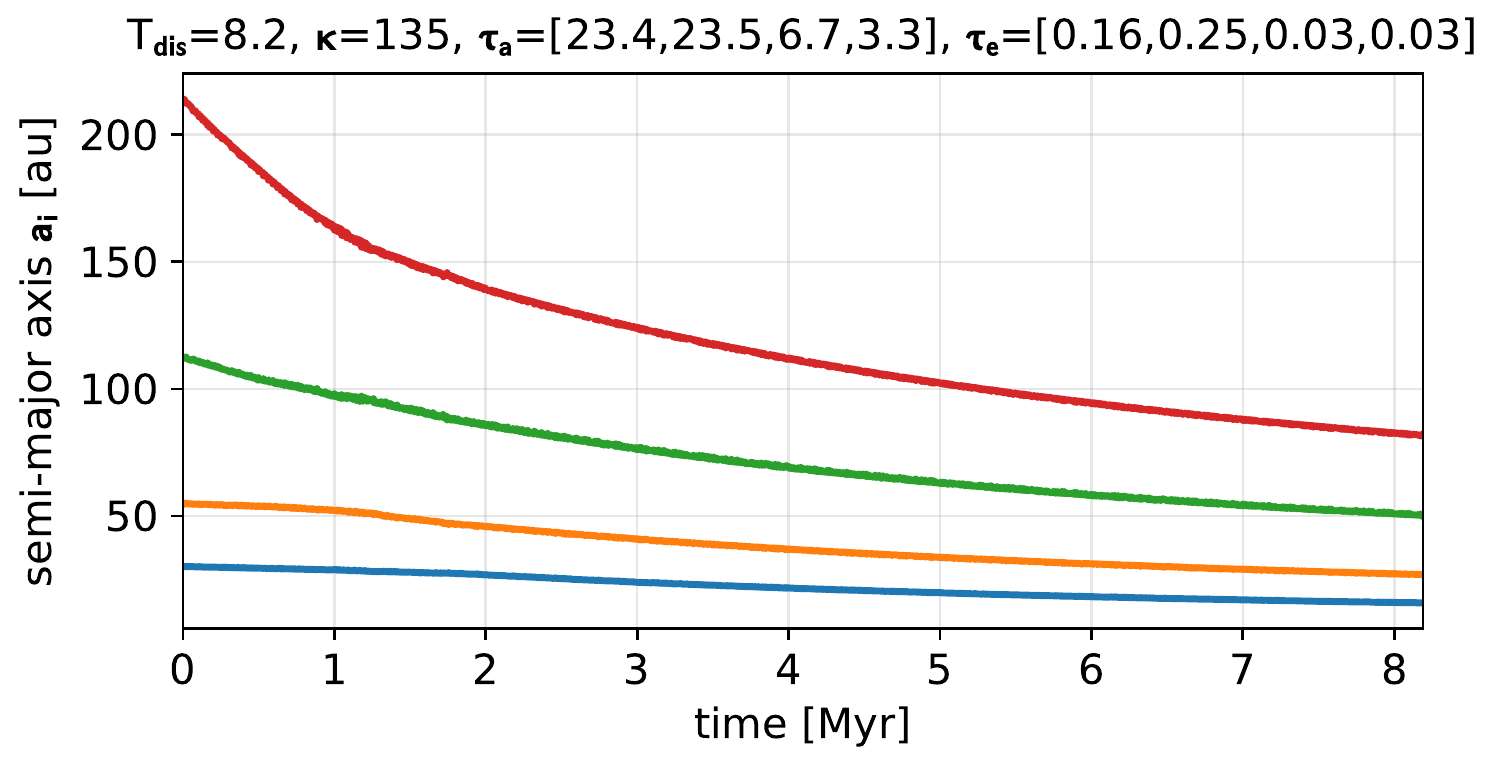}}
\hbox{\includegraphics[height=0.16\textheight]{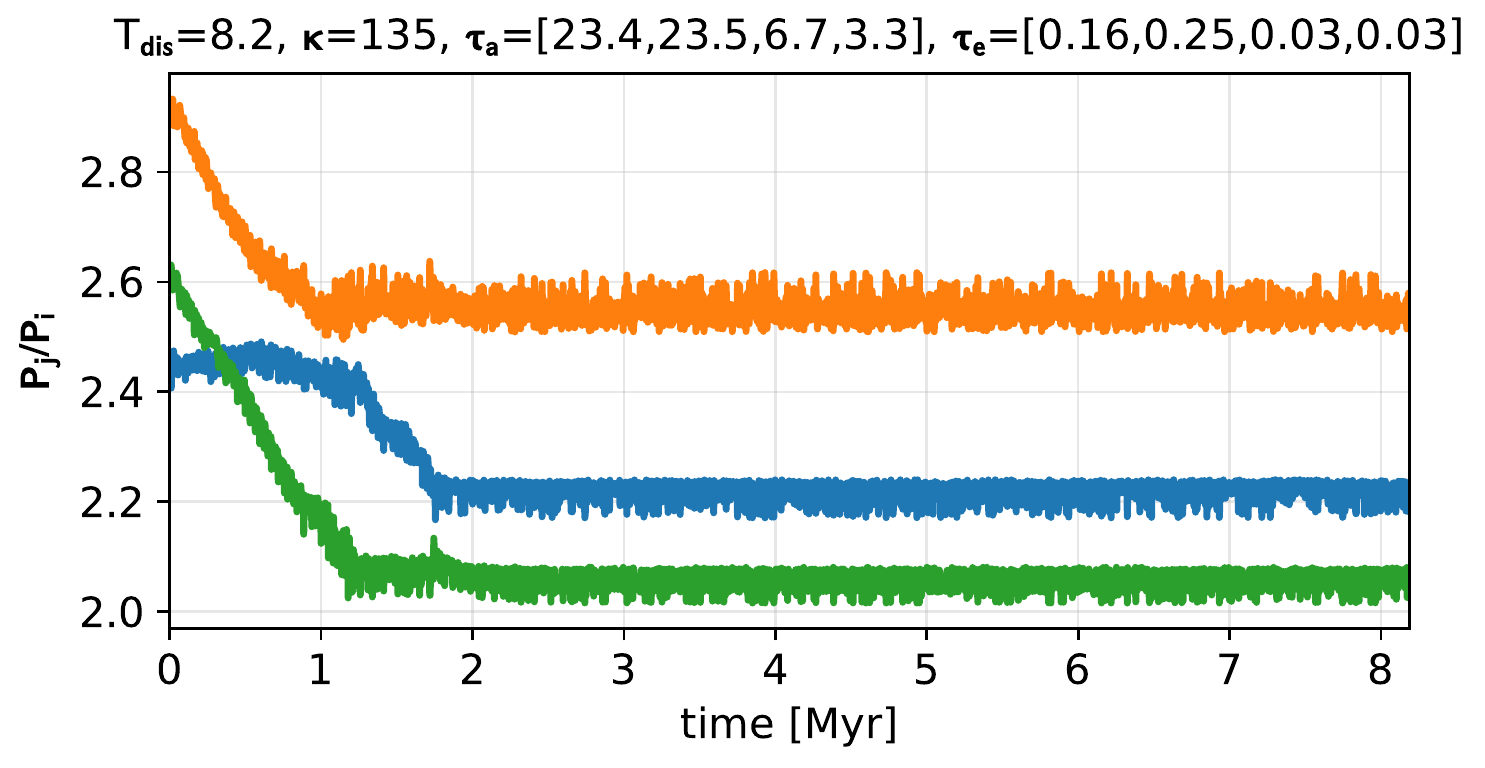}}
\hbox{\includegraphics[height=0.16\textheight]{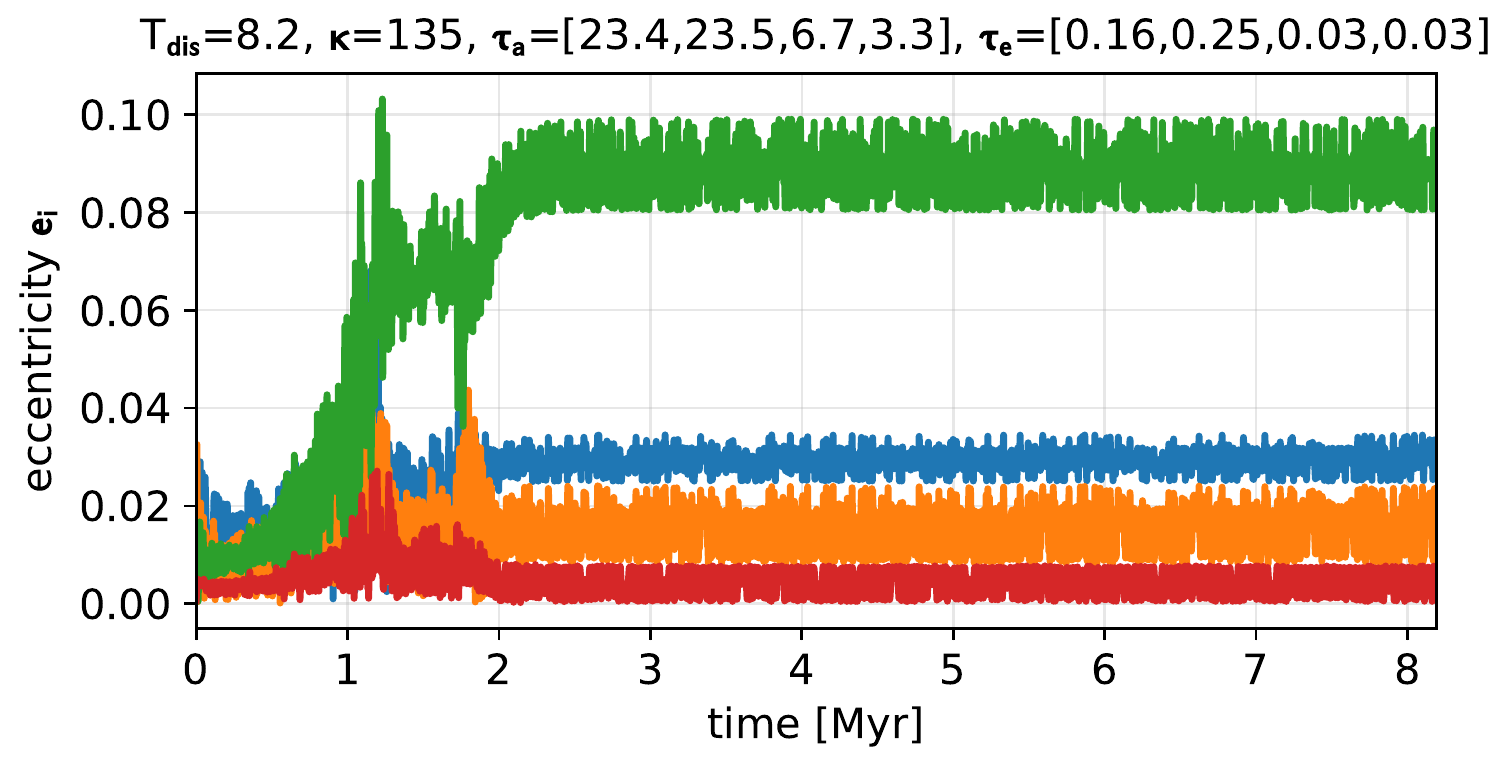}}
\hbox{\includegraphics[height=0.16\textheight]{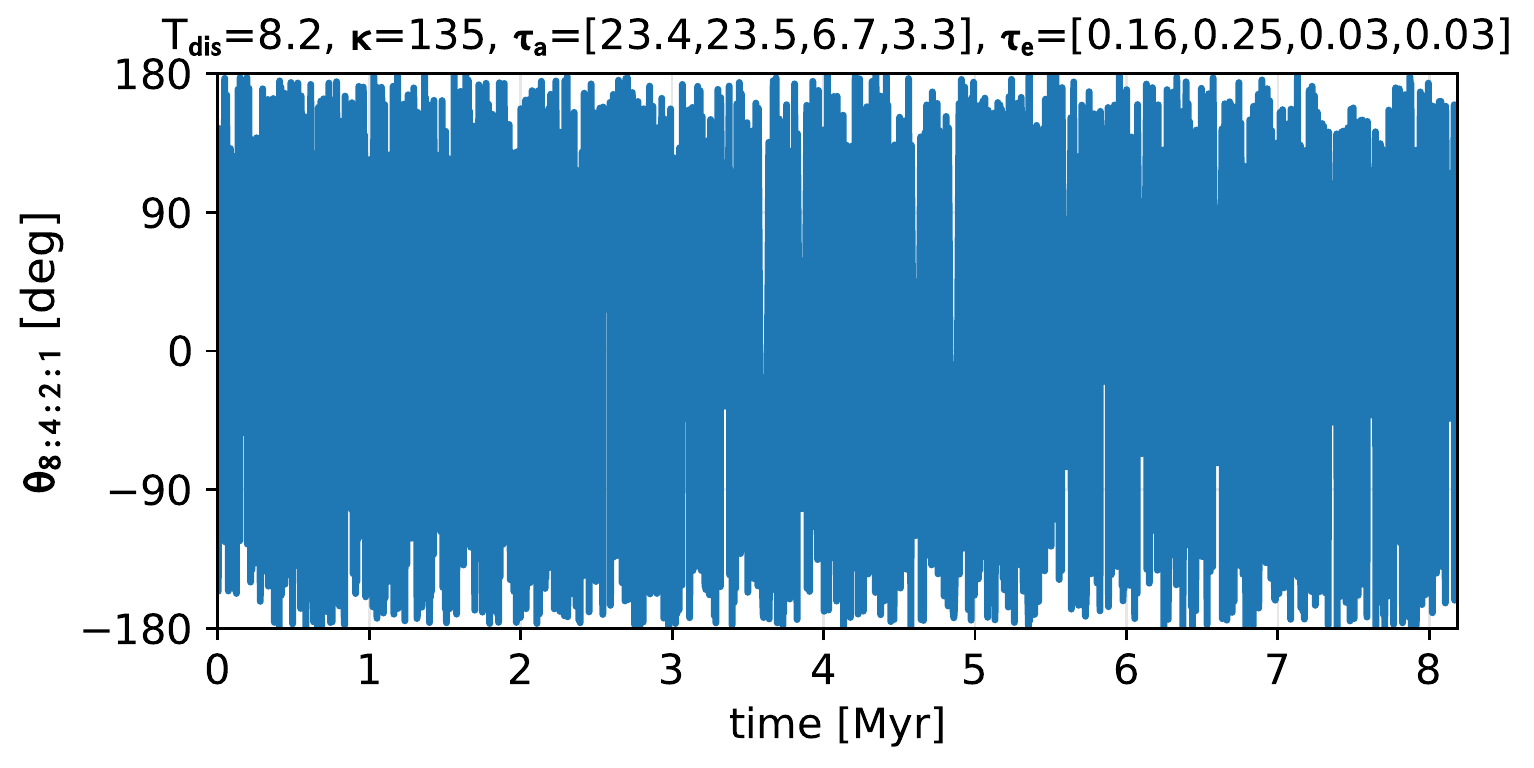}}
\hbox{\includegraphics[height=0.16\textheight]{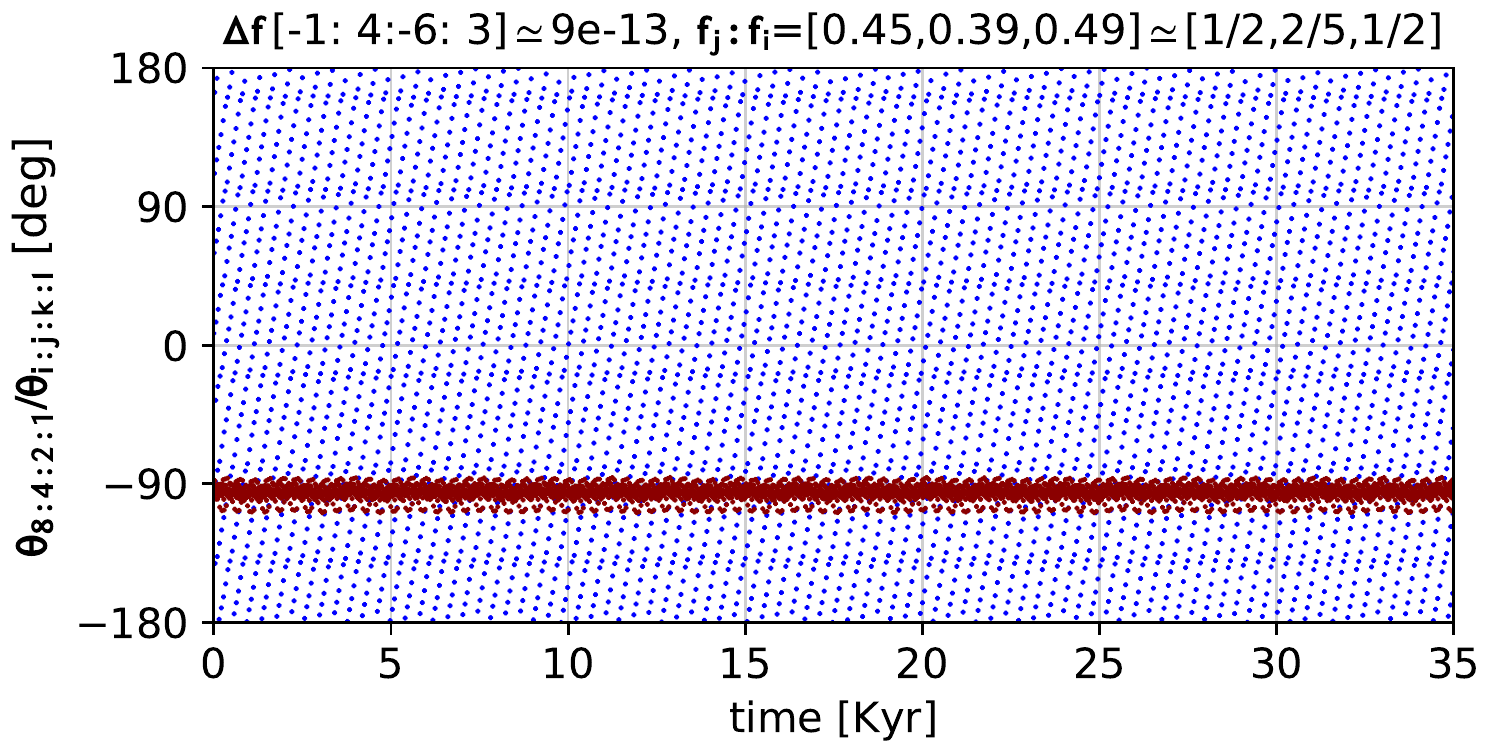}}
}
}
}
\caption{
The migration capture of four \hr8799{} planets into the generalized Laplace resonances. Panels from the top to bottom illustrate temporal evolution of the semi-major axes, orbital period ratios of subsequent pairs of planets, eccentricities, {and the critical arguments ({\em two bottom panels}), respectively.  The disk decay timescale $T_{\rm dis}$, as well the migration time-scales $\tau_{\rm a}$ and $\tau_{\rm e} \equiv \kappa^{-1} \tau_a$ in square brackets are in Myrs; $P_j:P_i$ and $f_j:f_i$ mean ratios of the astrocentric orbital periods and the proper mean motions $f_i$ for subsequent pairs of planets; $\Delta f[\eta_1,\eta_2,\eta_3,\eta_4]$ is for the absolute range of the linear combination of the frequencies $f_i$, and is expressed in radians per the so called {\em long day} of $2\pi\,$yr$^{-1} \simeq 58.13$~d. All panels were generated in course of the simulations. See the text for more details}. 
}
\label{fig:fig2}
\end{figure*}

{
The migration experiments described below were conducted with the initial semi-major axes by {three-four} times as large as in the observed system, yet each one selected randomly within $50$\% deviation from its nominal values for the \LPR{} Laplace resonance. The initial eccentricities were selected randomly, $e_{1,2,3,4} \in [0,0.16)$. The pericenter arguments $\varpi_i$ and the mean anomalies ${\cal M}_i$ were drawn from the uniform distribution in $[0^{\circ},360^{\circ}$). As for the planetary masses $m_i$, we choose the uniform distribution limited within
$[6,{10}$]~$\mJ$ range. 
}

{
In order to introduce a variability in the outcome configurations, we considered a few variants of Eq.~\ref{eq:migration}, with $\kappa_i$ selected randomly for each planet around a~mean $\kappa \in [1,300)$, or by choosing it the same for all planets. We also randomly changed the dispersal time $T_{\rm dis} \in [1,30]$~Myr, or it was infinite. We randomised individual timescales of the migration $\tau_i \in [1,30]$~Myr, forming a decreased sequence, in order to obtain convergent migration.
}

{
The equations of motion were integrated until a traced system did not disrupt and the hierarchy of the initial semi-major axes was preserved. We stopped the integrations when the inner semi-major axis in the migrating systems becomes $a_1<14.6$~au. Then we integrated the $N$-body equations of motion without dissipation for additional 32768 steps of 400~days ($\sim 1/40$ of the innermost period), in order to determine the proper mean motions for all planets. The proper mean motions are the fundamental frequencies $f_i$ resolved with the refined Fourier frequency analysis
\citep{Laskar1993,Nesvorny1996} of the time series $\{a_i(t) \exp( \mbox{i} \lambda_i(t))\}$, where $a_i(t)$ and $\lambda_i(t)$ are {\em the canonical} osculating semi-major axis and the mean longitude, respectively, as inferred in the Jacobi or Poincar\'e frame. Then we computed the linear combinations of the fundamental frequencies,
\[
 \Delta f (\vec{n})\equiv \sum_{i=1}^N n_i f_i,
\]
where $\vec{n} \equiv [n_1,\ldots,n_N]$ is a vector of integers in the $[-6,6]$ range, which yield small $|\Delta f|$. In this way, we aimed to find possible, low-order multi-body mean motion resonances (MMRs chains). Simultaneously, we examined critical angles corresponding to $\vec{\eta} = \mbox{arg} \min |\Delta f (\vec{n})|$
\[
 \theta_{\vec{\eta}}(t) \equiv \sum_{i=1}^N \eta_i \lambda_i.
\]
If $|\vec{\eta}|=0$, in accord with the d'Alambert rule, then the resonant configuration may be called the generalized Laplace resonance \citep{Papaloizou2015}, or the multiple MMR chain of the zero-th order. 
}

{
Two typical examples of the resonance capture of four-planet systems are illustrated in Fig.~\ref{fig:fig2}. {\em The left column} is for a system trapped in the generalised Laplace resonance which exhibits librations of at least two critical angles, with  amplitudes of several degrees: the ``classic'' Laplace resonance, which we studied in
\citep{Gozdziewski2014A}, is distinguished through
\[
 \theta_{8:4:2:1} \equiv \theta_{1:-2:-1:2} = \lambda_1 -2 \lambda_2 -\lambda_3 +2\lambda_4,
\]
as well as another critical angle $\theta_{-1:4,-5:2}$. In this four-body MMR chain, the orbital period ratios are pairwise close to 2.
}

{
The second example in {\em the right column} of Fig.~\ref{fig:fig2} shows a zero-th order four-body MMR with librations of only one critical angle $\theta_{-1,4,-6,3}$ ($\theta_{8:4:2:1}$ circulates). Moreover, the middle pair of planets exhibits the orbital period ratio close to 2:5 (see {\em the bottom plot} with $\Delta f$ in Fig.~\ref{fig:fig2}).
}

{In both cases, the inner eccentricities are excited to moderate values of $\simeq 0.1$, which usually leads to good orbital fits.}

\begin{figure*}
\centerline{ 
\hbox{
\includegraphics[width=0.33\textwidth]{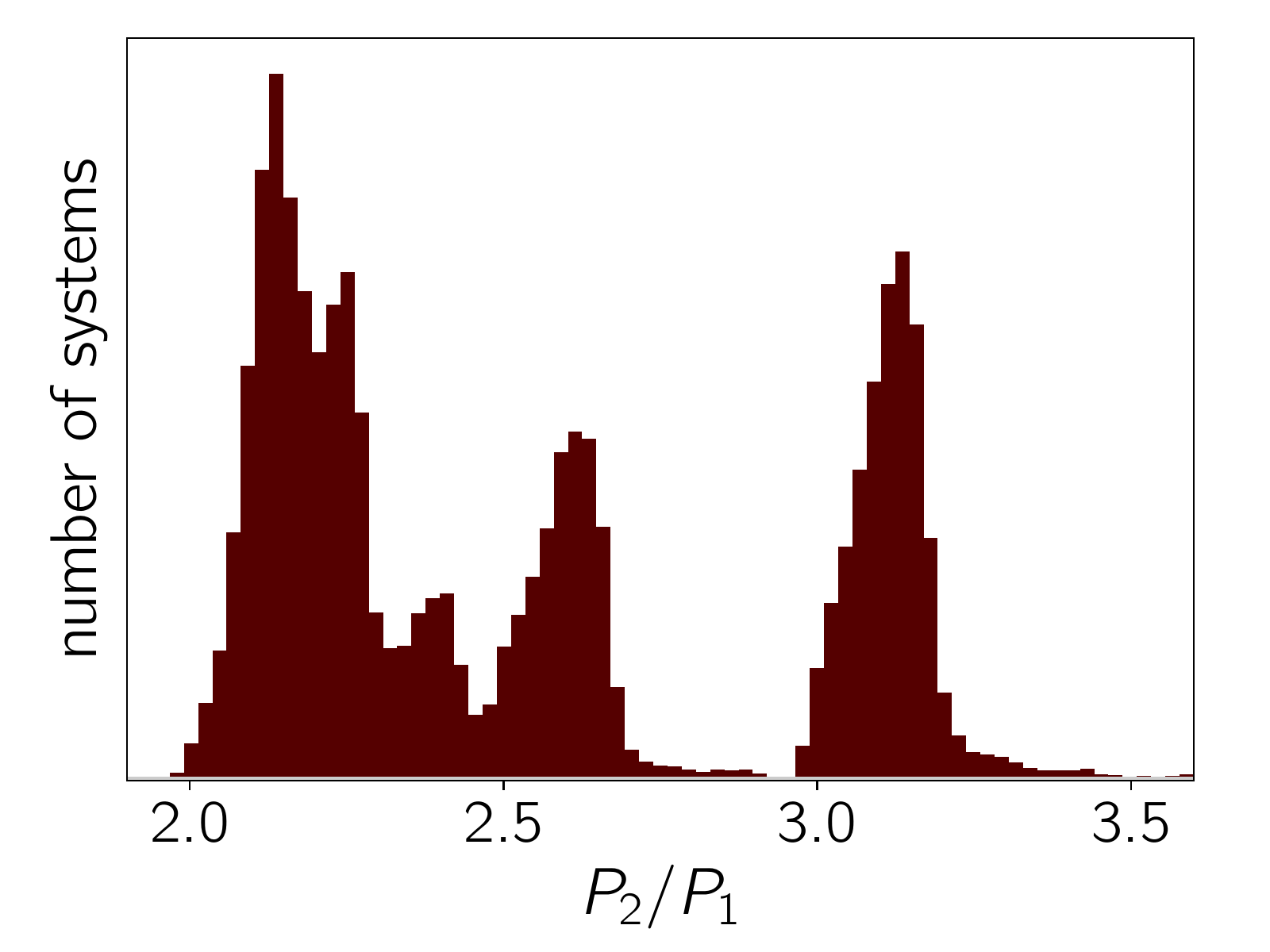}
\includegraphics[width=0.33\textwidth]{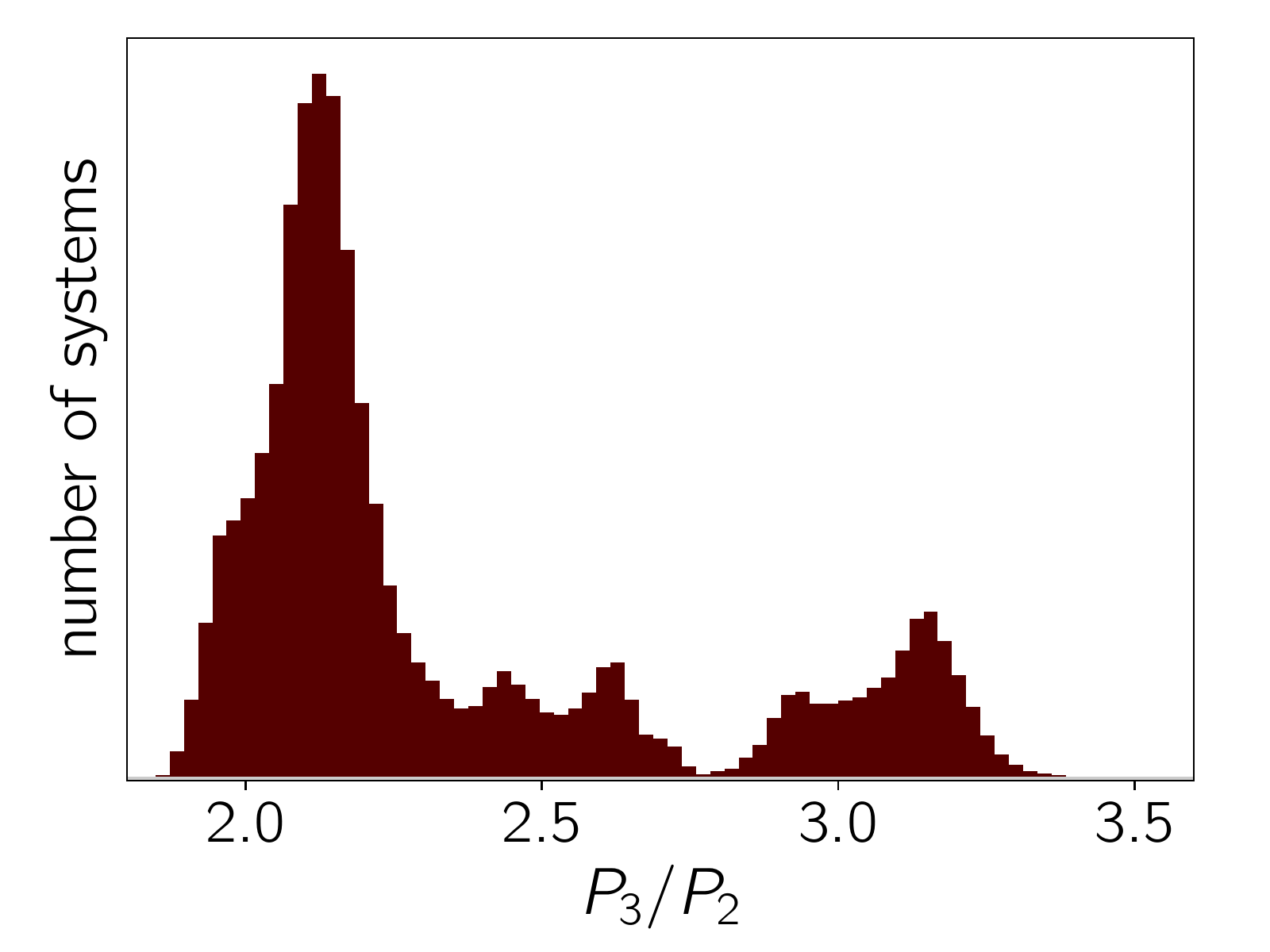}
\includegraphics[width=0.33\textwidth]{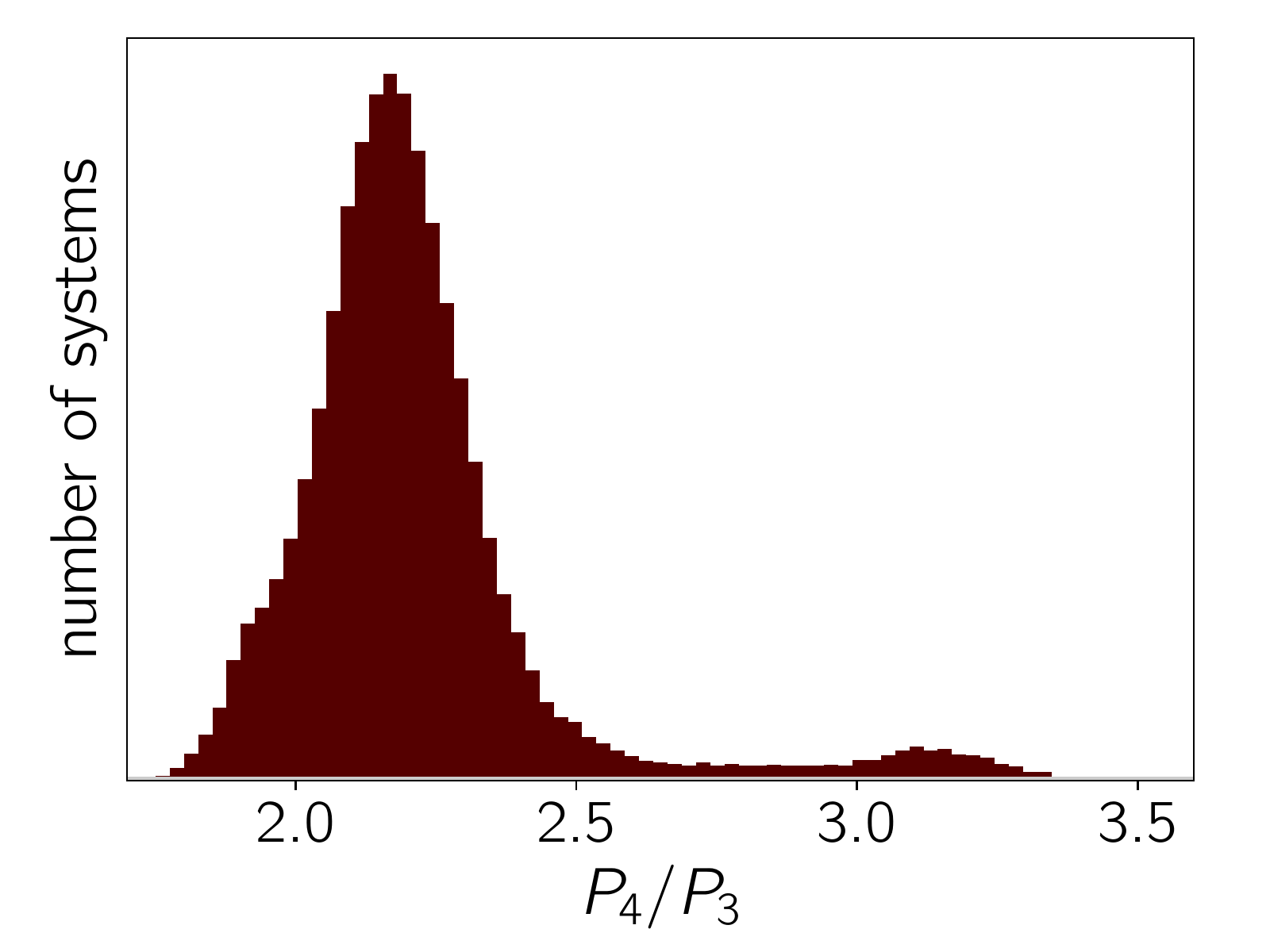}
}
}
\centerline{
\hbox{
\includegraphics[width=0.24\textwidth]{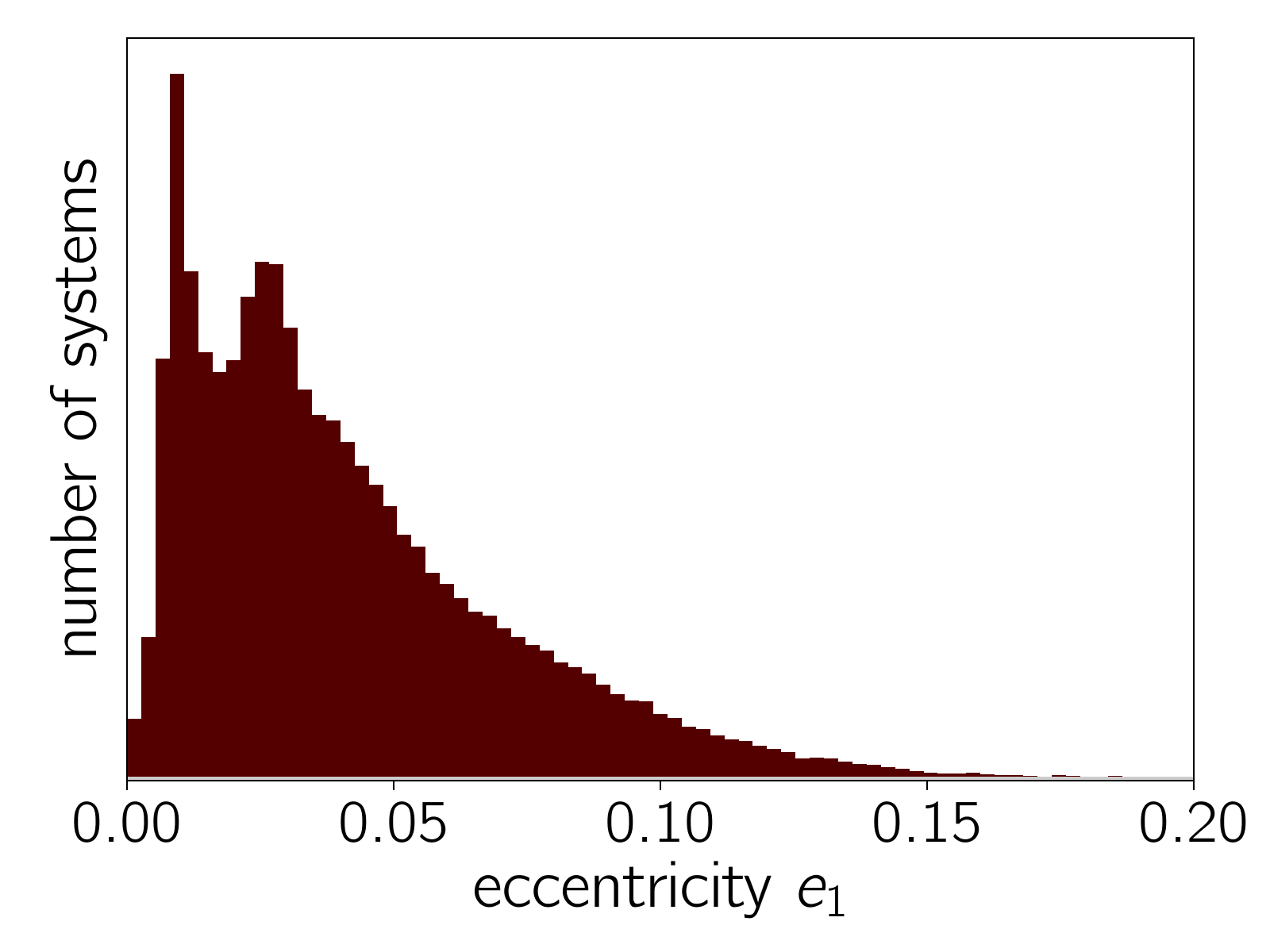}
\includegraphics[width=0.24\textwidth]{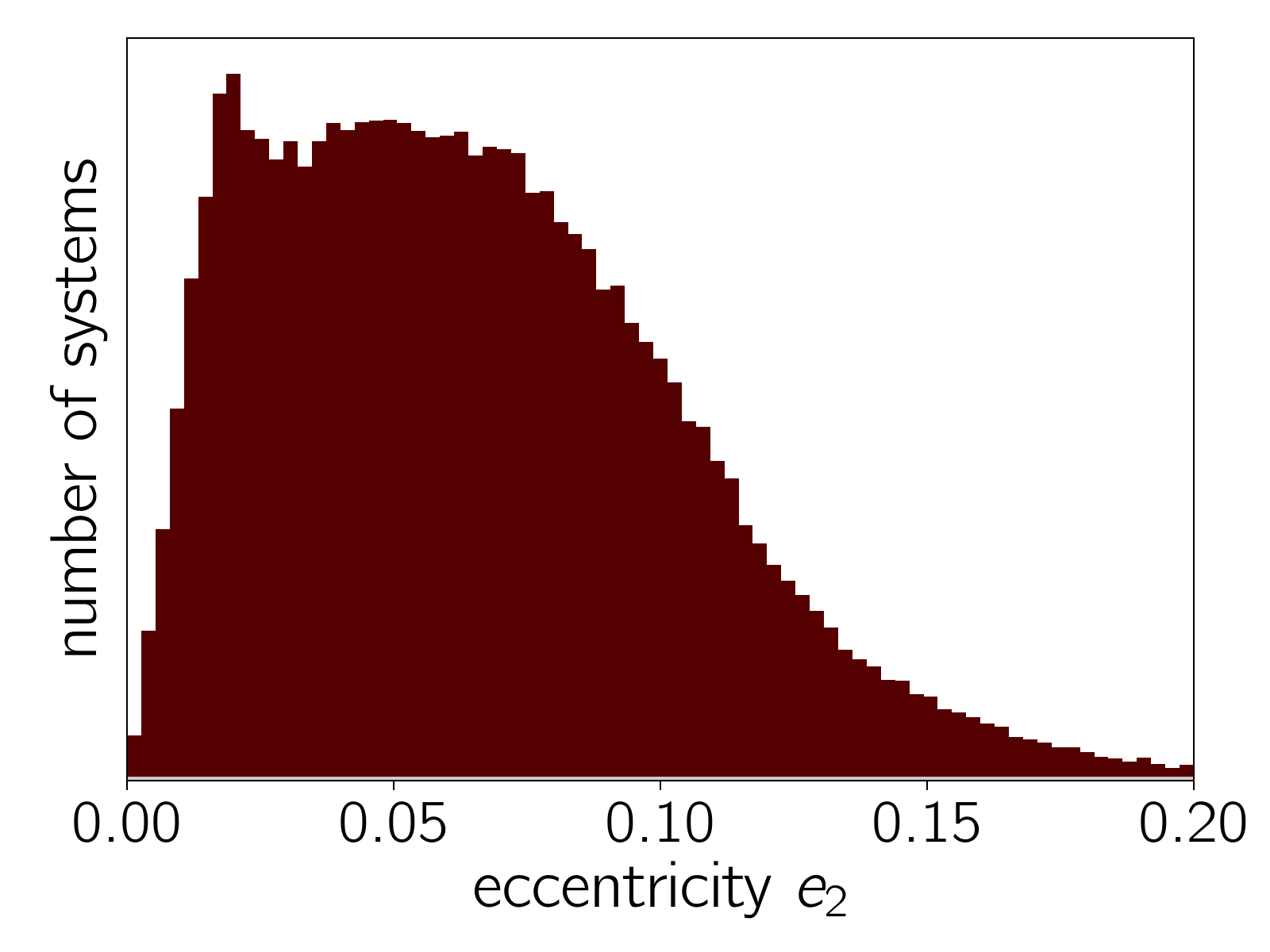}
\includegraphics[width=0.24\textwidth]{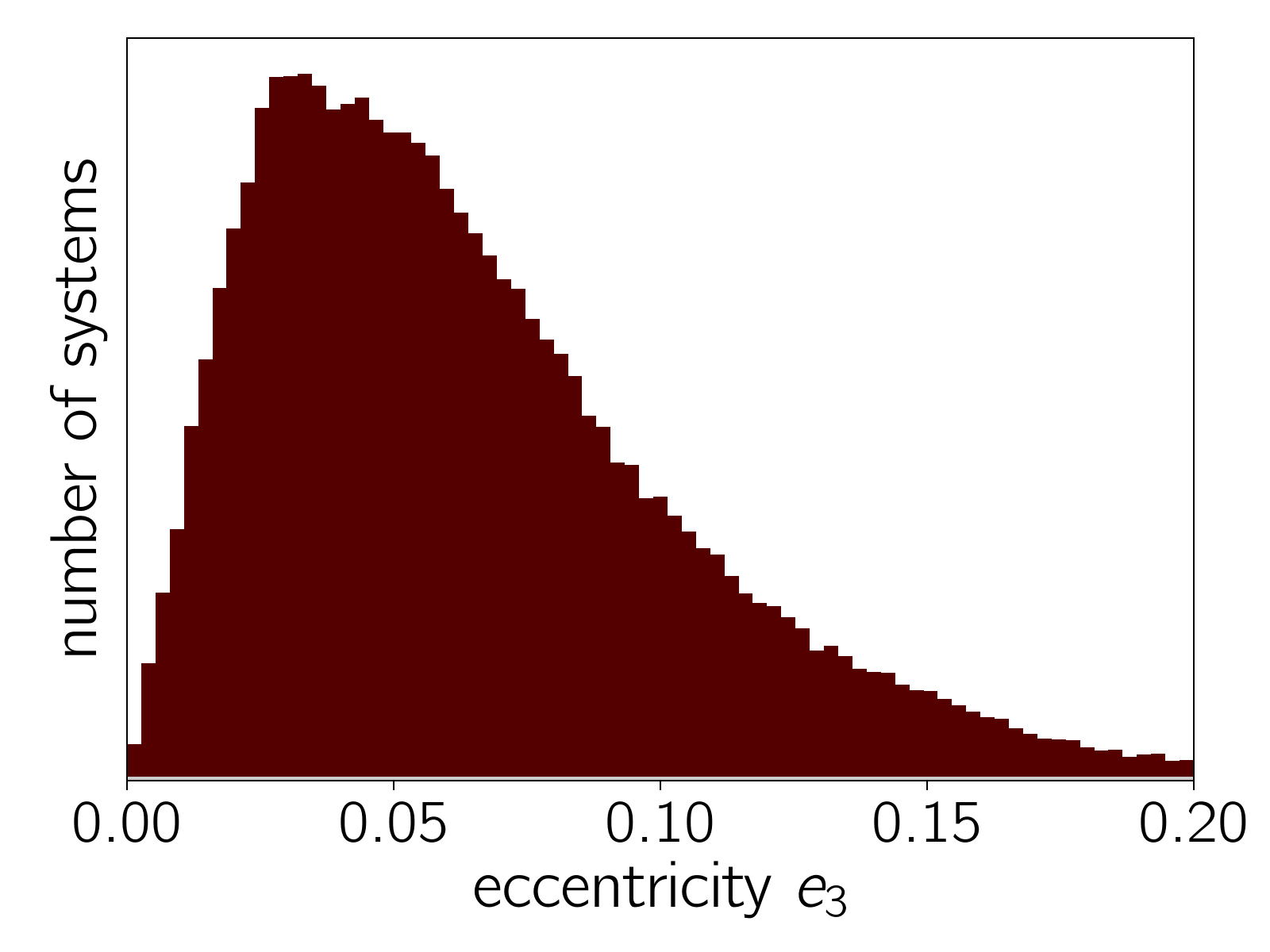}
\includegraphics[width=0.24\textwidth]{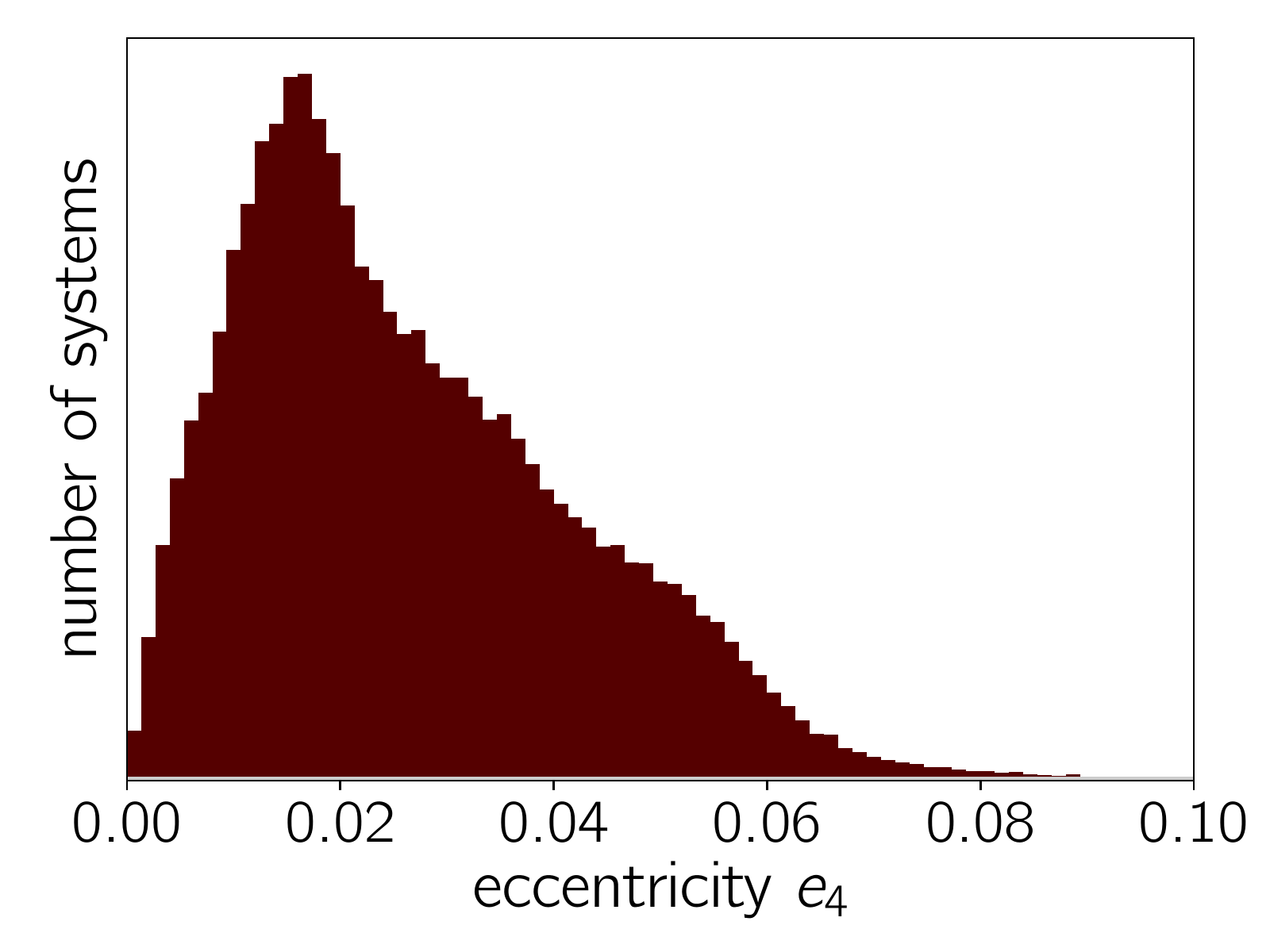}
}
}
\caption{
A graphical representation of the \moa{} database that consists of $\sim 1.5 \times 10^5$ four-planet configurations obtained through the planetary migration. {\em The top row} is for the distribution of osculating, astrocentric period ratios at the epoch zero, when inner semi-major axis $a_{\rm e}<14.6$~au. {Peaks in {\em the top row} indicate low-order two-body  MMRs}. {\em The bottom row} is for the distribution of osculating, astrocentric eccentricities at the zero-epoch.
}
\label{fig:fig3}
\end{figure*}

Figure~\ref{fig:fig3} shows the distribution of osculating astrocentric orbital periods and eccentricities attained when $a_1 < 14.6$~au.  At this moment, which is the reference zero-epoch for further optimization, the migration simulation was stopped. A feature of this distribution is an over-populated regime of 2:1 MMR for subsequent pairs of planets. The 2:1 MMR appears most ``easily'', in spite of significantly varied initial orbits, masses and migration parameters. {The final eccentricities may be as large as 0.2, yet the most frequent values are found around 0.02--0.05.}

{
The statistics of migrated configurations illustrated in Fig.~\ref{fig:fig3} involves also a significant fraction of systems with the innermost and the outermost pairs in the 3:1~MMR. However, such systems do not fit observations since the semi-major axis of \hr8799b{} appears too wide, $a_{\rm b} \sim 80$~au.
}

{An interesting result flowing from identification of the MMRs is a dominant proportion of the zero-th order MMR chains, which may be estimated as large as $\sim 90\%$ in the total sample of $\sim 1.5 \times 10^5$ systems. It may deserve further study, which we aim to conduct in a~separate work}.

{Similar results were obtained at early preparation of this paper, regarding five planets systems, with a hypothetical planet \hr8799{}f beyond the orbit of \hr8799{}b, and with a small mass of $\simeq 2~\mJ$ below the present detection level. In this case, the initial spread of semi-major axes in migrating systems was more close ($\pm 20\%$) to the 2:1~MMRs for subsequent pairs. Regarding eccentricity, the only qualitative difference with the four-planet simulations is a longer tail in the histogram for planet~b, extended for $e_{\rm b}\sim 0.15$, which is forced by the outer planet \hr8799{}f.}

Here, we consider essentially the most simple model of the migration. One could apply more sophisticated theories accounting for the mass growth, or stochastic migration \citep{Papaloizou2006,Armitage2018}. At this stage it is more important to gather a large set of orbital elements representing long-term stable systems, trapped in possibly different MMRs, {rather than analysing the process in fully realistic settings.} 

\subsection{Constraining MMR orbits with astrometric data}
At the next step~II, the MMR trapped systems from the \moa{} database are fine-tuned to fit particular or all available observations. Only a fraction of them may reproduce the observed system. Here, we rely on the scale-free property of the $N$-body equations of motion, which allows for scaling the semi-major axes through a factor of $\rho_a>0$, without changing a dynamical character of the scaled system.

We find the osculating epoch $t_0 \in [-T_0,T_0]$, relative to the end-time of the migration (zero-epoch), where $T_0$ is typically $\sim 10^3$ outermost periods, since the orbits have different orbital periods and they quickly precess. We also need to fit three Euler angles $(I,\Omega,\omega)$ which rotate the orbital plane of the original system to the observer (sky) plane.  The synthetic astrocentric signal is derived by propagating linearly re-scaled orbits through the numerical solution of the $N$-body equations of motion for time $t_0$, relative to the zero-epoch of the migration, with the initial eccentricities and orbital phases of their self-consistent, MMR-fixed values.  The parameters
$\vec{p}=(\rho_a,t_0,I,\Omega,\omega)$ are arguments of the merit function expressed as
$\Chi(\vec{p})$ or the maximum likelihood function $\log {\cal L}(\vec{p}) \sim
-\Chi(\vec{p})/2$ \citep{Bevington2003}.
We use the Bulirsh-Stoer-Gragg (BSG) numerical integrator to solve the equations of motion \citep{Hairer2001}. The absolute and relative local error limits are set to $\sim 10^{-15}$.

We underline that the astrocentric model of the observations is based on the self-consistent, canonical $N$-body dynamics, unlike Keplerian, geometric parametrization used frequently in the literature. The $N$-body model is more CPU-demanding, but it explicitly accounts for the planetary masses and the mutual gravitational interactions between the planets. We demonstrate its importance, when discussing the results (Sect.~\ref{sec:best}).

The optimization is performed for a number of MMR systems with the help of evolutionary algorithms \citep{Price2005,Izzo2012,Charbonneau1995}. Furthermore, the Bayesian inference and MCMC sampling makes it possible to introduce prior information on the system parameters, in order to estimate their uncertainties and correlations. We choose models yielding reasonably small $\Chi$ as the best-fitting solutions.

What separates the planetary migration (Step I) from the orbital optimization (Step II) is the scale-free property of the $N$-body Newtonian dynamics. In the barycenter frame they read as follows
\begin{equation}
\label{eq:nbody}
\ddot{\vec{r}}_i = k^2 \sum_j m_j \frac{\vec{r}_{ji}}{|{\vec{r}}_{ji}|^3}, 
\quad i\neq j=0,\ldots,N,
\end{equation}
where ${\vec{r}_{ji}} \equiv {\vec{r}}_j - {\vec{r}}_i$ is the relative radius vector from a body $i$ to a body $j$, where $i,j=0$ denotes the star, $m_j$ are for the masses ($m_0 \equiv m_{\star}$), and $k^2$ is the gravitational constant. The scaling invariance of the particular Ordinary Differential Equations (ODEs), here Eqs.~\ref{eq:nbody}, means that if a particular
$\vec{r}_{i}(t)$ is the solution to these equations, then $\rho_a^{-2/3}
\vec{r}_i(\rho_a t)$ with some fixed and constant scaling factor $\rho_a>0$ is also the solution. Therefore the orbital radii, or orbital arcs at prescribed time intervals,  may be scaled by the same factor, yet with an appropriate time change. Such a geometrically re-scaled copy of the orbits exhibits the same dynamical character as the original system.

\begin{figure*}
\centerline{ 
\vbox{
\hbox{\includegraphics[width=0.32\textwidth]{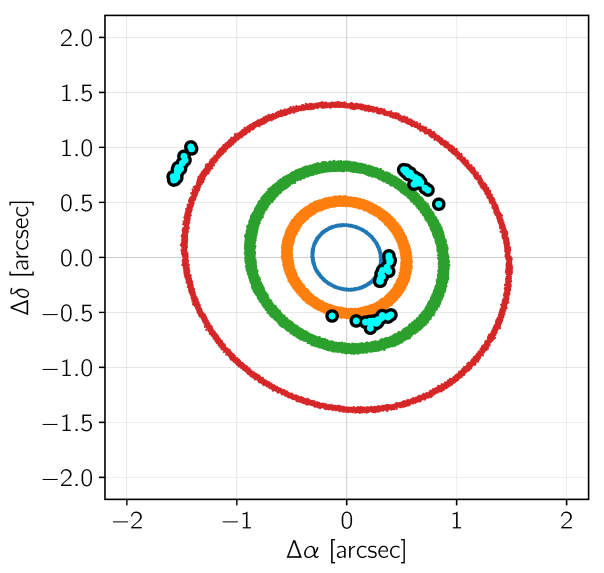}}
\hbox{\includegraphics[width=0.32\textwidth]{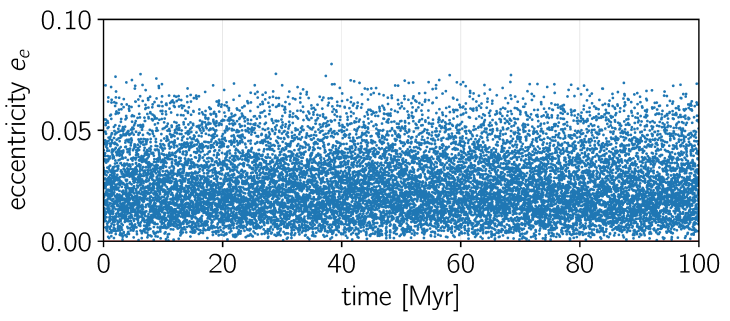}}
\hbox{\includegraphics[width=0.32\textwidth]{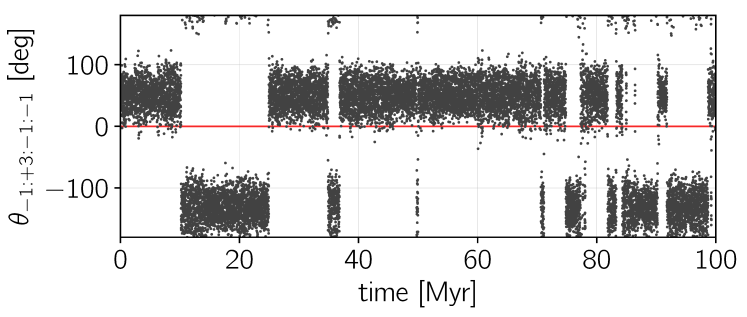}}
\hbox{\includegraphics[width=0.32\textwidth]{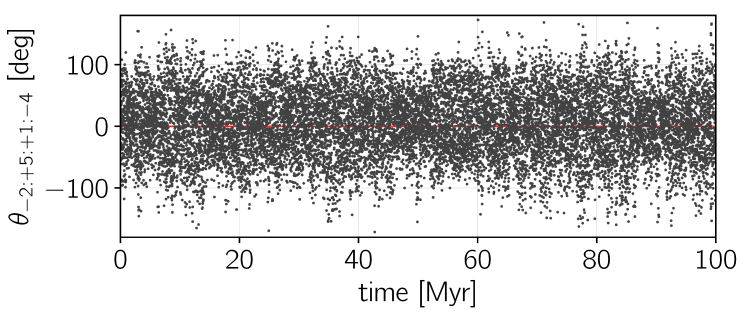}}
}
\vbox{
\hbox{\includegraphics[width=0.32\textwidth]{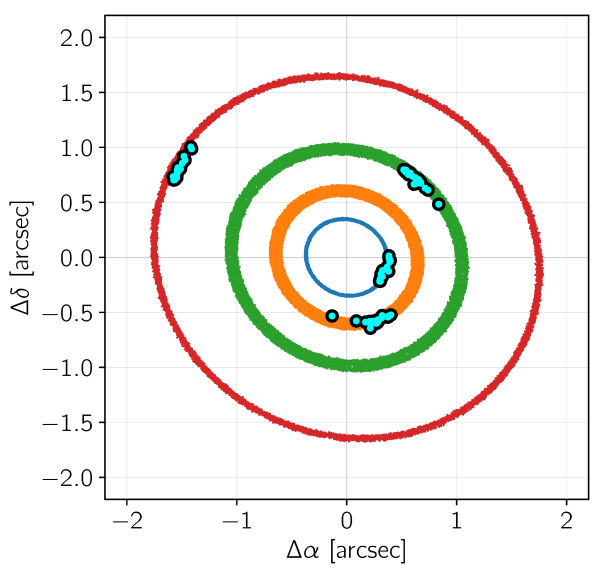}}
\hbox{\includegraphics[width=0.32\textwidth]{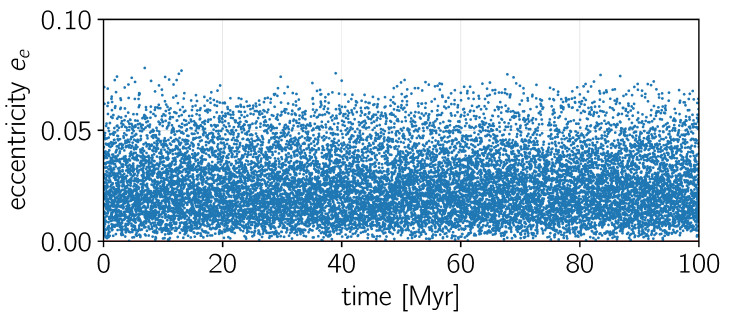}}
\hbox{\includegraphics[width=0.32\textwidth]{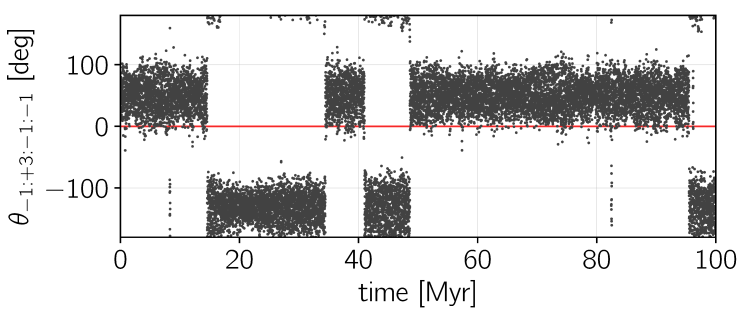}}
\hbox{\includegraphics[width=0.32\textwidth]{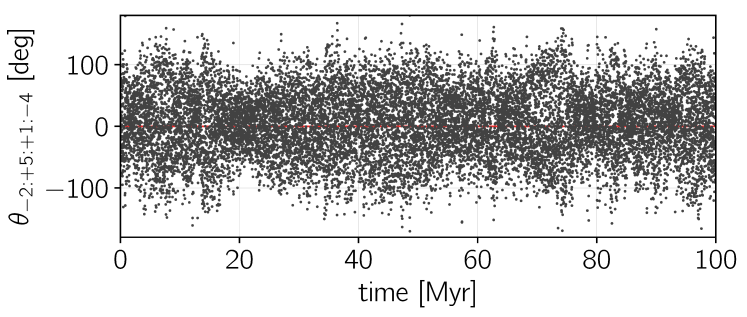}}
}
\vbox{
\hbox{\includegraphics[width=0.32\textwidth]{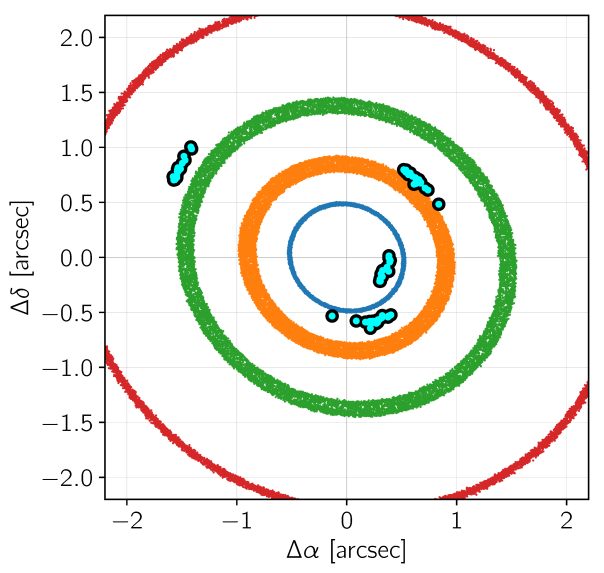}}
\hbox{\includegraphics[width=0.32\textwidth]{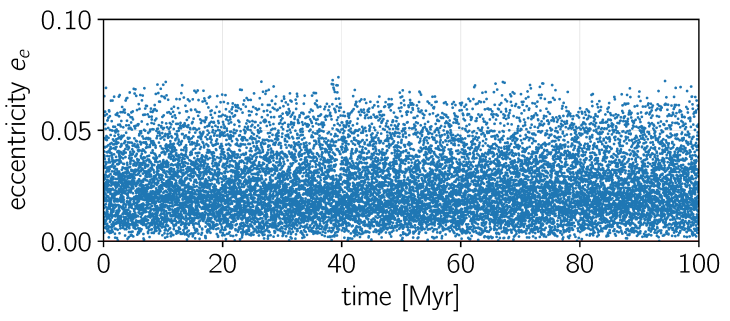}}
\hbox{\includegraphics[width=0.32\textwidth]{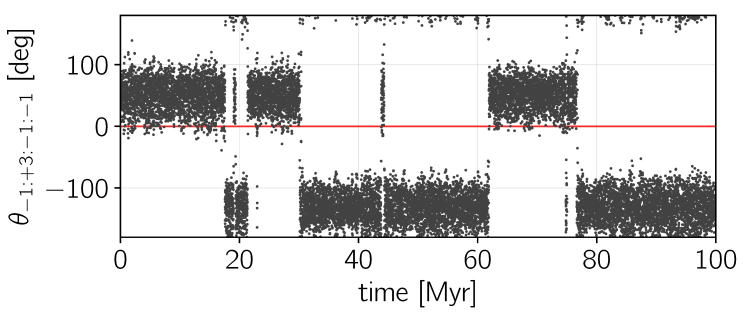}}
\hbox{\includegraphics[width=0.32\textwidth]{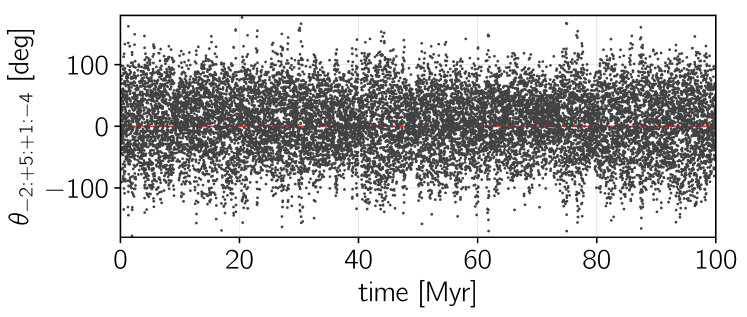}}
}
}
\caption{
Illustration of scale-free dynamics at the second step of \moa{}. Initial semi-major axes in a resonant configuration of four planets shown in the {\em middle column} are linearly scaled by a factor of $0.8$ ({\em the left column}), and by a factor of $1.33$ ({\em the right column}), respectively. We selected two critical arguments of the {generalized} Laplace resonance, $\theta=\lambda_1-3\lambda_2+\lambda_3+\lambda_4$ and $\theta=2\lambda_1-5\lambda_2-\lambda_3+4\lambda_4$, respectively ({\em two middle rows}) which librate with similar amplitudes ({\em two bottom rows}). The innermost eccentricity ({\em the second row from top}), shown as an example, varies within similar limits.
}
\label{fig:fig4}
\end{figure*}

The $N$-body ODEs scaling invariance is illustrated in Fig.~\ref{fig:fig4}.  By following a migrating planetary configuration, we usually end up with too compact ({\em the left column}) or too wide ({\em the right column}) orbits, with respect to the observations. In fact, these configurations are re-scaled copies, by a factor of 0.8 and 1.33, respectively, of the directly simulated system in the {\em middle column}. Apparently, it fits relatively well to the observations. This resonant configuration has been obtained through a fast migration during only { 1.8}~Myr. In spite of a substantial scaling, the dynamical character of all configurations is preserved for at least 100~Myr: the critical angles of the zero-th order {generalized} Laplace resonance \citep{Papaloizou2015} oscillate around the same center, as well as the eccentricities vary within the same limits ({\em the middle-} and {\em bottom} panels in Fig.~\ref{fig:fig4}). 

We note that the reference configuration is weakly chaotic in the sense of a small mLCE. However, in this Lagrange-stable configuration, all orbits remain bounded for a very long time. The system remains stable since it is resonant, actually in an unusual way -- one of the critical arguments librates around the same libration center, while the second one ``switches'' between two centers. This example is selected intentionally, to show that even weakly chaotic configurations may be scaled without loosing any of the geometric features, like the critical angles or eccentricity evolution. 

The scale-free dynamics property releases us from an uncertainty of the star distance. The present determination of the parallax $\pi=24.22 \pm 0.09$ mas ($\sim 41.3$~pc) in the GAIA Data Release~2 catalogue \citep{Brown2018} places the system almost 2~pc farther than assumed in the literature to date, $d=39.4$~pc \citep[][]{Leeuwen2005}. The relatively significant correction of the parallax has essentially no implication on the orbital models, besides the planetary masses may be larger due to a correction for the absolute luminosity.

\section{The best-fitting orbital model}
\label{sec:best}
We searched for the best-fitting solutions to the astrometric measurements in \citep{Konopacky2016} with the approach described above.  Homogeneous observations of the HR 8799 system in \citep{Konopacky2016} are obtained between 2007 and 2014 with NIRC2 on the Keck~II telescope and with the same reduction pipeline. The data are corrected for systematic biases present in observations in the discovery papers \citep{Marois2008,Marois2010}. Keplerian (kinematic) solutions in \citep{Konopacky2016} favour co-planar, low eccentric orbits, as well as agree to within $1\sigma$ with the
\LPR{} resonance model, hence our assumptions are supported by their independent analysis.

We aim to verify whether the \moa{} model of these observations, which are a subset of data available to date and gathered in \cite[][their Appendix]{Wertz2017}, may be extrapolated and fit all measurements made by different authors with other instruments. We also found a significant deviation of the synthetic model orbit IVa in \citep{Gozdziewski2014A} for the outermost planet, which systematically precedes the latest observations in \citep{Pueyo2015} and \citep{Maire2015}. The new analysis could improve the previous model. 

Among the found solutions, we selected the best-fitting configuration yielding $\Chi~\simeq 0.87$. Its osculating elements at epoch $t_0=\epk{}$ of the first observation in \citep{Konopacky2016}, as determined from the primary fit parameters
$\vec{p}$,  are displayed in Tab.~\ref{tab:tab1} (model IV$_{\rm K}$). We found tens of geometrically similar configurations within $\Chi\simeq 1$, therefore we focus on this particular, basis configuration. This model, representing the \LPR{} MMR chain, is similar to the best-fitting solution IVa in \citep{Gozdziewski2014A}. Yet we note a smaller value of the initial semi-major axis $a_4 \simeq 67$~au. We examined this difference closely. As expected, the new solution improves the astrometric model by providing a better match to the first and the last observations at epochs of 1998.83 and 2014.93 in the whole data set, respectively. Indeed, it removes the systematic trend and the ``too fast'' orbital motion of planet~b predicted by our earlier model IVa.

\begin{table*}
\caption{
Osculating elements of the best-fitting four- and five-planet solutions. For model IVa reproduced from \citep{Gozdziewski2014A}, the osculating epoch is $1998.83$, and the stellar mass $m_{\star} = 1.56\,\msun$. Model IV$_{\rm K}$ found in this work is determined at the osculating epoch of \epk{} and for the star mass of $m_{\star} = 1.52\,\msun$. Co-planar solutions postscripted with ``A'', ``B'', ``C'', ``D'' and ``E'' extend model IV$_{\rm K}$ with a fifth hypothetical planet in a stable orbit. 
Model IV$_{\rm MCMC}$ is a non-coplanar four-planet stable solution with small $\Chi \simeq 1$ found with the MCMC sampling around model IV$_{\rm K}$.
{These models are solutions to the measurements in \citep{Konopacky2016}.}
Model V$_{\rm A}$ is a preliminary fit to all observations collected in \citep{Wertz2017}, and found with the \moa{} for a five-planet system.
Uncertainties for parameters in models IV$_{\rm K}$ and IV$_{\rm MCMC}$ are determined as ranges of $\Ym$-stable samples yielding $\Chi<1.6$ around their median values. Formal errors, determined as the 16-th and 86-th percentile of the samples are by 2--3 times smaller. For a reference, parameter values for fits IV$_{\rm K}$ and IV$_{\rm A}$ with five digits after decimal point are provided, as literally used in our experiments.
Parameter $\nu=N_{(\alpha,\delta)} - \dim{\vec{p}}$ is for the degrees of freedom, where $N_{(\alpha,\delta)}$ denotes the number of $(\alpha,\delta)$ measurements, and
$\dim{\vec{p}}$ is the number of free parameters.
}  
\label{tab:tab1}
\centering
\begin{tabular}{l c c c c c c c}
\hline
\hline
& $m\,[\mJ]$ & $a\,$[au] & $e$ & $I\,$[deg] & $\Omega\,$[deg] & $\varpi\,$[deg] & $\Mmean\,$[deg] \\
\hline
\hline
\multicolumn{8}{c}{ 
Model IVa at epoch of $1998.83$, $m_{\star} = 1.56\,\msun$ (Fig.~\ref{fig:fig7}, {\em top row})} \\
\hline
\hr8799{}e & $9 \pm 2$ & $15.4 \pm 0.2$ & $0.13 \pm 0.03$ & & & $176 \pm 6$ & $326 \pm 5$\\
\hr8799{}d & $9 \pm 3$ & $25.4 \pm 0.3$ & $0.12 \pm 0.02$ & $25 \pm 3$ & $64 \pm 3$ & $91 \pm 3$ & $58 \pm 3$\\
\hr8799{}c & $9 \pm 3$ & $39.4 \pm 0.3$ & $0.05 \pm 0.02$ & &  & $151 \pm 6$ & $148 \pm 6$\\
\hr8799{}b & $7 \pm 2$ & $69.1 \pm 0.2$ & $0.020 \pm 0.003$ & & & $95 \pm 10$ & $321 \pm 10$\\
\hline
\hline
\multicolumn{8}{c}{ 
Model IV$_{\rm K}$ at epoch of \epk{},  $m_{\star} = 1.52\,\msun$, $\Chi=0.87$, $\nu=97$,
$\dim{\vec{p}}=5$  (Figs.~\ref{fig:fig8}--\ref{fig:fig11})}\\
\hline
\hline
\hr8799{}e & $9.4 \pm 0.5$ & $15.45 \pm 0.25$ & $0.127 \pm 0.011$  & & & $107.5 \pm 1.5$ &
${13.7} \pm 1.5$\\
 &  9.426       &  15.45001  & 0.12696 & & & 107.54637   &   13.67259 \\
\hr8799{}d & $8.2 \pm 0.8$ & $25.36 \pm 0.30$ & $0.095 \pm 0.006$  & & & $26.9 \pm 2.5$ & $79.4 \pm 3.0$\\
& 8.185 & 25.35505   &    0.09540 & & &  26.94575   &   79.41486 \\
\hr8799{}c & $6.9 \pm 0.5$ & $39.78 \pm 0.40$ & $0.048 \pm 0.005$  & $25 \pm 2$ & $64 \pm 5$ &
${105.8} \pm 2.0$ & ${137.2}\pm 5.0$\\
&6.857  &  39.77699    &   0.04829 & 25.289  &  64.414 &  {105.79908}   &
{137.18943} \\
\hr8799{}b & ${6.7} \pm 0.5$ & $67.01 \pm 0.35$ & $0.023 \pm 0.003$ & & & $
{120.1} \pm 3.0$ & ${235.7} \pm 5.0$\\
&6.680 & 67.00881  &     0.02297 & & &  {120.07543}  &  {235.66545} \\
\hline
\hr8799{}fA & 1.660 & {115.25600} & 0.02222 & & & 174.59807 & 54.13554 \\ 
\hr8799{}fB & 0.660 & 116.43105 & 0.02222 & & & 174.59807 & 54.13554 \\
\hr8799{}fC & 1.000 & 134.16610 & 0.02218 & 25.289 & 64.414 & 11.33805 & 327.42380 \\ 
\hr8799{}fD & 0.330 & 133.86900 & 0.02218 & & & 11.33805 & 327.42380 \\
\hr8799{}fE & 0.100 & 137.76200 & 0.02218 & & & 11.33805 & 327.42380 \\
\hline
\hline
\multicolumn{8}{c}{ 
Model IV$_{\rm MCMC}$ at epoch of \epk{}, $m_{\star} = 1.52\,\msun$, $\Chi=0.98$, $\nu=74$, $\dim{\vec{p}}=28$} (Fig.~\ref{fig:fig6}) \\
\hline
\hline
\hr8799{}e & $9.5 \pm 0.5$ & $15.48 \pm 0.25$ & $0.123 \pm 0.011$  & $26 \pm 2 $ & $64 \pm 5 $ & $110.2 \pm 1.5$ & $ 12.5 \pm 1.5$ \\
& 9.489 & 15.48223  &   0.12278  &  26.138  & 63.612 &  110.23990 &  12.49904 \\
\hr8799{}d & $7.5 \pm 0.8$ & $25.35 \pm 0.30$ & ${0.097} \pm 0.006$  & $29 \pm 2$ & $56 \pm 5 $ & $ 33.1 \pm 2.5$ & $ 80.8 \pm 3.0$ \\
& 7.490 & 25.35133  &   0.09669  &  29.096  & 56.160  &  33.05848 &  80.78110 \\
\hr8799{}c & $7.1 \pm 0.5$ & $39.95 \pm 0.40$ & $0.046 \pm 0.005$  & $25 \pm 2 $ & $63 \pm 5 $ & $104.8 \pm 2.0$ & $140.0 \pm 5.0$ \\
& 7.082 &   39.94936 &    0.04629  & 25.318 &   62.732 & 104.77760  & 139.98150 \\
\hr8799{}b & $6.8 \pm 0.5$ & $67.11 \pm 0.35$ & $0.024 \pm 0.003$  & $31 \pm 2 $ & $60 \pm 5 $ & $118.5 \pm 3.0$ & $242.9 \pm 4.5$ \\
& 6.753 &   67.11155 &   0.02358  & 30.517 &  59.595 & 118.49210 & 242.91630 \\
\hline
\hline
\multicolumn{8}{c}{ 
Model V$_{\rm A}$ at epoch of 1998.83,  $m_{\star} = 1.52\,\msun$, $\Chi=2.71$, $\nu=229$,
$\dim{\vec{p}}=5$ (Fig.~\ref{fig:fig7}, {\em bottom row})} \\
\hline
\hline
\hr8799{}e & $9.450$ &  15.45863  & 0.09789 &     &  & 115.18157 & 330.89613 \\
\hr8799{}d & $6.851$ &  25.59693  & 0.09300 &     &  &  29.67592 &  62.45914 \\
\hr8799{}c & $7.121$ &  39.78074  & 0.05227 & 26.103 &  60.210 & 101.42421  &  136.56084  \\
\hr8799{}b & $8.847$ &  70.24633  & 0.05055 &   &    &  52.58777 & 306.83016 \\
\hr8799{}f & $2.500$ & 111.48326  & 0.01205 &   &    & 140.41787 &  65.69502 \\
\hline
\hline
\end{tabular}
\end{table*}
 

\subsection{Measuring stability of orbital solutions}
\label{sec:Ym}

In order to reveal the global dynamical structure of the system, we use the fast indicator technique, besides the direct integration of the equations of motion. The idea behind this approach relies in determining the character of motion (chaotic or regular) on relatively short orbital arcs. A configuration classified as the fast indicator-stable for relatively short motion time of  $\simeq 10^4$ outer periods may be extrapolated for 10-100 longer Lagrange stability time (called also the event time, $T_{\rm E}$), implying non-crossing, non-colliding and bounded orbits.  Usually, the fast indicators are  related either to the maximal Lyapunov Characteristic Exponent \citep[mLCE,][]{Cincotta2000,Gozdziewski2008} or to a diffusion of the fundamental frequencies
\citep{Laskar1993}.

As the fast indicator, essentially equivalent to the mLCE, we use the Mean Exponential Growth factor of Nearby Orbits (MEGNO, or $\Ym$ from hereafter) developed by \cite{Cincotta2000,Cincotta2003}. It is implemented in our Message Passing Interface (MPI) parallelized code \mufarm{}. The required system of the equations of motion and their variation equations may be integrated with the Bulirsh-Stoer-Gragg (BSG) scheme \citep{Hairer2001}. We also used the symplectic algorithm \citep{Gozdziewski2008}, which is much more CPU efficient, but it might be non-reliable in strongly chaotic zones where collisional events are possible. We decided to use the BSG scheme for final experiments, to avoid such numerical and artificial biases.

For brevity, orbital configurations {are called} $\Ym$-stable if $\mmegno<\epsilon$, where $\epsilon\simeq0.05$, for a particular number of characteristic periods, counted in $10^4$ outermost planet orbits. The MEGNO integrations were stopped if $\mmegno>5$, and we consider such configurations as unstable ($\Ym$-unstable). As we found in previous papers, $\Ym$-stable models are equivalent to Lagrange-stable solutions for one-two orders of magnitude longer intervals of time.

More details on these MEGNO calibration to determine the Lagrange stability, regarding the \hr8799{} system, and also low-mass planet systems were discussed in our earlier papers \citep{Gozdziewski2008,Gozdziewski2014A} as well as in a new work \citep{Panichi2017}. We also note that \megno{} has been recently used as a reference tool by \cite{Hadden2018}, who developed a new, quasi-analytic stability criterion for eccentric two-planet systems.

\subsection{The MCMC experiments setup}

We conducted followup MCMC experiments, aiming to determine the orbital parameter uncertainties for stable solutions found with \moa{}, and to demonstrate the degree of instability of the four-planet system.  For that purpose we performed single-temperature MCMC sampling with the \emcee{} package \citep{ForemanMackey2013}. Recalling our results in \citep{Gozdziewski2014A}, the orbital elements in stable four-planet models may be only varied within tiny ranges. Here, we release the co-planarity constraint and we search for stable solutions around the best-fitting model IV$_{\rm K}$. The Bayesian inference makes it possible to introduce prior information, like indirect observational constraints. For instance, the debris disks geometry determined independently of the imaging astrometry are reported almost coplanar with the planetary orbits, {and seen at the position angle around of $50^{\circ}$--$60^{\circ}$ \citep{Matthews2014,Booth2016}. Therefore the debris disk models could impose additional, indirect constraints} on the inclination $I$ and nodal angle $\Omega$ of the system. 

With the MCMC sampling of the parameter space, stable planetary configurations might be detected, perhaps relatively distant from the model IV$_{\rm K}$ in Tab.~\ref{tab:tab1}. Therefore, {in the search for stable solutions}, we conducted the MCMC sampling with up to 2048 \emcee{} walkers initiated in a small hyper-cube in the orbital parameter space centered at the best-fitting model, for 128,000 to 256,000 iterations each, {mostly limited by CPU-demanding stability checks made with the MEGNO indicator}. We evaluated $\Chi$ and the likelihood function $\log {\cal L} \sim -\Chi/2$ required to compute the posterior distribution. In this experiment, masses of the planets and all orbital elements are considered as free parameters of the astrometric model.

In order to narrow the parameter space, {and to reduce possible degeneracies}, we imposed Gaussian priors ${\cal N}(\mu,\sigma_{\mu})$ with the mean $\mu$ equal to the best-fitting values in Tab.~\ref{tab:tab1}. As for the variances $\sigma_{\mu}$, we choose: for the masses, $\sigma_{m}=3\,\mJ$; for the semi-major axes, $\sigma_{a}=5$~au; for the Poincar\'e elements $x_i=e_i \cos\omega_i$, and $y_i=e_i \sin\omega_i$,  $\mu_{x,y}=0$ and $\sigma_{x,y}=0.3$; for the inclinations and the nodal longitudes,  $\sigma_{I}=30^{\circ}$ and $\sigma_{\Omega}=60^{\circ}$, respectively. The priors have been set as improper (uniform) for the osculating mean anomalies at the initial epoch, ${\cal M}_i \in[ 0^{\circ},360^{\circ})$, $i=1,\ldots,4$. 

During the MCMC sampling with the Gaussian priors, we also evaluated the MEGNO of all solutions with $\Chi<2.6$, to determine limits of stability zone around the best-fitting stable solution. The $\Ym$ integration time was set to $\simeq 7,000$ outermost periods, hence permitting for marginally Lagrange-stable models. In accord with our earlier experiments \citep{Gozdziewski2014A}, the $\Ym$ integration interval should be safely longer than $10^4$ characteristic periods, in order to determine long-term Lagrange-stable configurations for at least 160~Myr. The $\Ym$-stable solutions for $7,000$ outermost orbits should provide ten times longer Lagrange stability time (at least $\simeq 30$~Myr).

For a reference, we also performed the MCMC sampling with all priors uniform, and determined in wide parameter ranges, spanning 24 $\mJ$ for masses $m_i$, and 30~au for semi-major axes $a_i$ around their best-fitting values IV$_{\rm K}$ in Table~\ref{tab:tab1}, as well as $x_{i},y_{i} \in [0,0.67]$ for the Poincar\'e elements, $I_i \in [0^{\circ},90^{\circ}]$, $\Omega_i\in [0^{\circ},180^{\circ}]$, and ${\cal M}_i \in [0^{\circ},360^{\circ}]$ ($i=1,\ldots,4$).

{
We tried to estimate the auto-correlation time $\tau_{\rm emcee}$ through sampling experiments without stability checks, in order to reduce the CPU overhead. We increased the chain length up to $512,000$. We used a method of \cite{Sokal1996}, as proposed by \citep{ForemanMackey2013} in the recent version of the \emcee{} sampler. We found that $\tau_{\rm emcee}$ is typically very long and varies between $\sim 40,000$ and $\sim 120,000$ for different elements and priors. The second parameter expressing the sampling ``health'', the acceptance rate, was typically well below 0.2, unless the \emcee{} scaling parameter $a_{\rm emcee}$ was set to low values of $\sim 1.2$. Similarly difficult and ill-conditioned Monte-Carlo and MCMC sampling experiments are reported by \cite{Konopacky2016} and \cite{Wertz2017}, regarding kinematic (Keplerian) models. }

{We found a particularly strange discrepancy of the distribution of the node arguments $\Omega_i$. As reported in \cite{Konopacky2016}, two $\Omega_i$ peaks wide for up to a few tens degrees are centered around roughly $\sim 60^{\circ}$ and $\sim 130^{\circ}$--$150^{\circ}$, respectively, for all planets but \hr8799{}d with one, much wider dominant peak around $\Omega_{\rm d} \sim 100^{\circ}$.  The first mode around $\sim 60^{\circ}$ is consistent with the outer  debris disk orientation \citep{Matthews2014,Booth2016}. However, it is missing at all  among single-mode posteriors of $\Omega_{\rm c}$ and $\Omega_{\rm e}$ derived by \cite{Wertz2017}. We note here that our $N$-body MCMC experiments made with fixed masses in the best-fitting solution IV$_{\rm K}$, that could mimic the Keplerian model, yet with Gaussian priors on the eccentricity, reveal a two-modal posterior only in $\Omega_{\rm c}$, while all remaining peaks of $\Omega_{\rm b,d,e}$ are found between $50^{\circ}$ and $60^{\circ}$. This most strong two-modal distribution of $\Omega_{\rm c}$ divided with a shallow posterior valley might explain a low acceptance rate in our experiments. Also a dynamical sense of the second mode of $\Omega_i \simeq 130^{\circ}$, implying mutually inclined orbits, remains uncertain.
}

\subsection{The MCMC experiments and the best-fitting model}

\begin{figure*}
\centerline{ 
\hbox{
\includegraphics[width=0.24\textwidth]{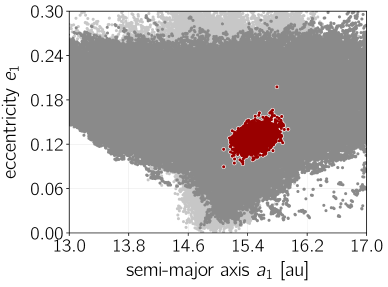}
\includegraphics[width=0.24\textwidth]{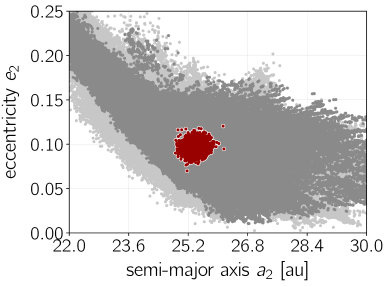}
\includegraphics[width=0.24\textwidth]{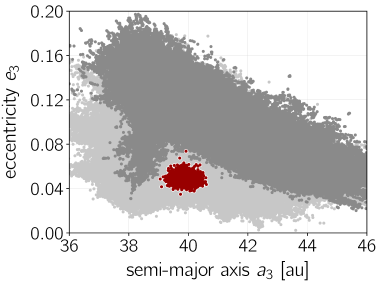}
\includegraphics[width=0.24\textwidth]{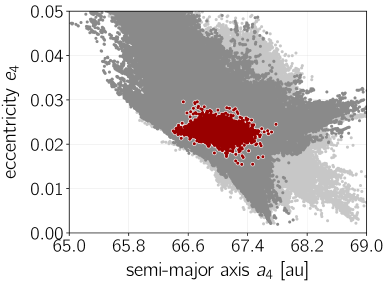}
}
}
\centerline{
\hbox{
\includegraphics[width=0.24\textwidth]{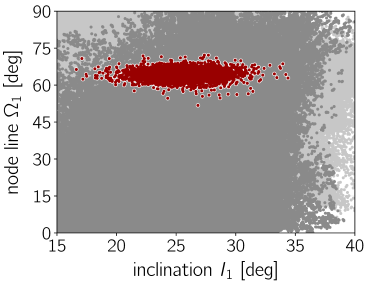}
\includegraphics[width=0.24\textwidth]{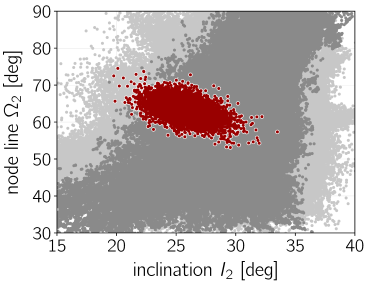}
\includegraphics[width=0.24\textwidth]{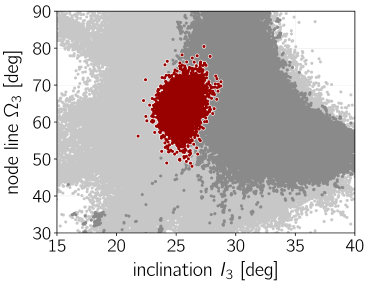}
\includegraphics[width=0.24\textwidth]{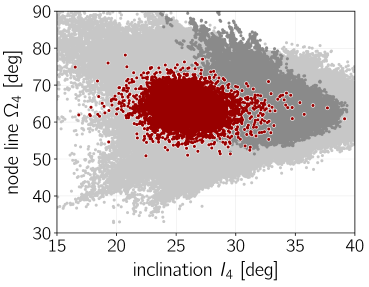}
}
}

\caption{
The results of MCMC experiments as 2-dim $\Chi$ projections of samples onto selected parameter planes. Light-gray and dark-grey filled circles mark solutions derived with uniform priors, with $\Chi<1.6$ and $\Chi<0.74$, respectively. Dark-red filled circles for $\Ym$-stable models with $\Chi<2.6$. See Fig.~\ref{fig:fig6} for the sky-plane representation of configurations selected from the illustrated samples.
}
\label{fig:fig5}
\end{figure*}

Our results derived in two example MCMC experiments are illustrated in Fig.~\ref{fig:fig5}, as 2-dim $\Chi$ distributions for selected parameters. The light-grey filled circles represent solutions with $\Chi<1.6$ and darker filled circles are for samples with $\Chi<0.74$, derived with uniform priors. We note that the lowest $\Chi\simeq 0.60$. The dark-red filled circles mark solutions $\Ym$-stable for $\sim 7,000$ outer periods when Gaussian priors are set only for masses and semi-major axes. The number of illustrated samples is $\sim 2 \times 10^6$.

Clearly, even for the very limited $\Chi<0.74$ range, the model parameters are practically unbounded. They vary in wide limits both for the uniform and Gaussian priors. Also the 1-dim posterior distributions (not shown) imply that the $N$-body  astrometric model is unconstrained, similarly to the Keplerian models in \citep{Konopacky2016} and \citep{Wertz2017}. Stable solutions are found only relatively close to the initial resonant configuration, and the overall shape of stable, compact zones agrees with the results derived with constrained genetic algorithm \citep[see,][for details]{Gozdziewski2014A}. Here, however, the search for stable non-coplanar solutions has been performed with {an independent method}. 

We performed the same experiment for different choices of the Gaussian priors (the mean values and their $\sigma_{\mu}$) for masses and orbital elements. All these attempts to find stable configurations resulted in outputs qualitatively similar to those demonstrated in Fig.~\ref{fig:fig5}. We estimate that the total number of tested samples with $\Chi<2.6$ has reached $10^9$ in more than 20 MCMC experiments performed on 256 CPU cores each. We did not find any $\Ym$-stable model beyond a close neighborhood of the best-fitting, stable resonant model~IV$_{\rm K}$. Therefore finding a stable configuration by chance, or without imposing tight or a'priopri constraints on the model parameters, would be extremely difficult.

\begin{figure*}
\centerline{ 
\vbox{
\hbox{
\includegraphics[width=0.33\textwidth]{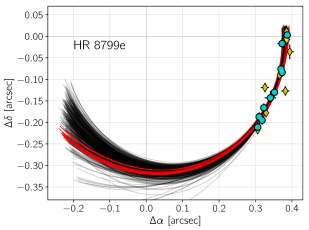}
\includegraphics[width=0.33\textwidth]{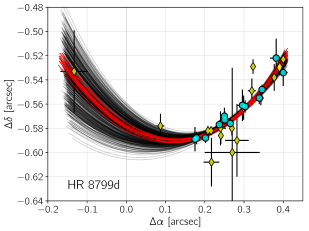}
\includegraphics[width=0.33\textwidth]{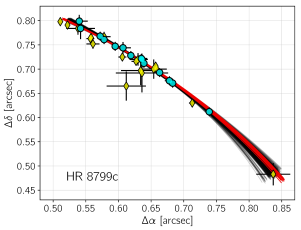}
}
\hbox{
\includegraphics[width=0.33\textwidth]{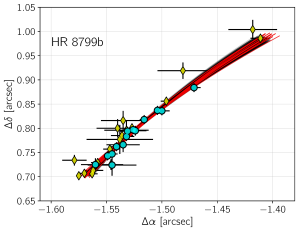}
\includegraphics[width=0.33\textwidth]{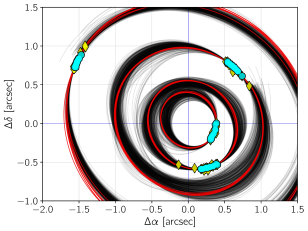}
\includegraphics[width=0.33\textwidth]{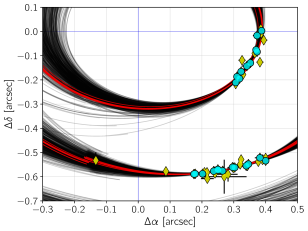}
}
}
}
\caption{
The best-fitting solution to the four-planet model illustrated at the sky-plane. The $y$-axis corresponds to N, and the $x$-axis corresponds to E direction, respectively (note that the numerical values of $\Delta\alpha$ are sign-opposite to regarding the formal left-hand direction of the right ascension $\alpha$). Light-blue (white-gray) filled circles are for measurements in \citep{Konopacky2016}, and dark-blue (dark-grey) filled circles are for measurements in other papers, as collected in \citep{Wertz2017}. Red curves mark several stable solutions with $\Chi \simeq 1$ from the MCMC sampling. Grey curves are for orbital arcs derived from unstable models within $\Chi<0.74$, marginally worse from $\Chi \simeq 0.6$ for the mathematically best-fitting $N$-body solutions. All model orbits have been derived for measurements in \citep{Konopacky2016}, and then extended between the epochs of 1998.83 and 2014.93 in the full data set in \citep{Wertz2017}. The osculating epoch is \epk{}, i.e., the first epoch in \citep{Konopacky2016}.
}
\label{fig:fig6}
\end{figure*}

Finally, the results of the MCMC sampling experiments shown in Fig.~\ref{fig:fig5} are further illustrated in Fig.~\ref{fig:fig6}. We selected 20~stable solutions with $\Chi<1.04$, and 500 other (unstable) models with low $\Chi<0.74$ from the MCMC-derived posterior samples. Their orbital arcs, marked with red and grey curves, respectively, are over-plotted on the astrometric data in \citep{Konopacky2016} as blue, filled circles, and all other measurements in the literature collected in \citep{Wertz2017}, with yellow diamonds. In spite of the tight restriction, $\Chi<0.74$, the grey curves span a wide region of the sky. Moreover, stable models to a subset of data in \citep{Konopacky2016} extrapolate very well to the full data set. We may note the proper timing of the synthetic orbits both with the most recent, as well as the earliest measurements ({\em the top-middle}, {\em the top-right} and {\em the bottom-left} panels in Fig.~\ref{fig:fig6}). 

An interesting conclusion regards {\em the top-middle} panel for planet \hr8799d{}. Low $\Chi<0.74$ solutions (dark curves) widely spread around the Hubble Space Telescope (HST) observation  at epoch 1998.98 in \citep{Lafreniere2009}. However,  stable, resonant models pass in the middle of these solution orbits, closely to the HST measurement made a few years before. It might be a lucky coincidence, but we recall that stable solutions {\em are extrapolated} back from the model epoch of \epk{}, and the HST measurement was not included in the optimization.

One more feature of stable solutions is illustrated in {\em the right-bottom} panel of Fig.~\ref{fig:fig6}, showing model orbits for the two inner planets. In this panel, each of the orbits are computed for their osculating period. Surprisingly, orbital arcs of stable solutions for planet HR~8799d {\em do not close}. It means that the gravitational interactions may be detected even within the narrow observational window. It is also a warning that kinematic models which do not account for the mutual planetary interactions {may be soon non-adequate in a longer time-basis}. 

{
In order to summarise, we consider the results of our MCMC experiments mostly as an attempt of determining the range of stable islands, rather than a rigorous optimisation and statistical inference presented in \citep{Pueyo2015,Konopacky2016,Wertz2017}. Moreover, since stability zones are extremely tiny with respect to wide posteriors, the results of statistically reliable MCMC sampling made with the kinematic orbital model (without stability checks) might bring only limited information on stable solutions. 
}

%
\section{The $\Ym$-model of debris disks}
%
\label{sec:calibration}
Having the long-term stable configurations of the system, we may simulate debris disks in a framework of the restricted and non-restricted problems. In the first version, mass-less particles (asteroids) are influenced by the gravitational tug of the planets (primaries), and we assume that the asteroids do not attract the planets nor mutually interact.  In the second, non-restricted case, hypothetical massive bodies may be added to the system of observed planets. Such bodies may be detected indirectly through resolving the global architecture of the observed system. For instance, a signature of the fifth planet beyond \hr8799{}b may be the radius of the inner edge of the outer debris disk \citep{Booth2016}. Also \cite{Wilner2018} constrained a mass of the outer planet \hr8799{}b to $\simeq 6 \mJ$ through modeling the inner edge with ALMA and VLA observations. Based on the ALMA observations alone, \cite{Read2018} deduced a mass of an additional planet \hr8799{}f beyond the orbit of planet~b. 

{We conducted extensive Monte Carlo simulations of the debris disks in both frameworks. The initial semi-major axis, as well as eccentricity and orbital phases are drawn randomly within prescribed ranges}. These elements extend the initial condition for the observed planets (primaries), and the resulting orbits are integrated. We then analyze a large volume of the initial conditions with two numerical methods.

For simulating the debris disk as the restricted problem, we use the direct $N$-body integrations with the hybrid scheme in the \mercury{} package \citep{Chambers1999}, corrected by \cite{Torres2008}. This approach is common in the literature \citep[e.g.,][]{Contro2016,Read2018}. We fixed the step-size of 64~days for the mixed variables leapfrog (MVS) and the local accuracy $\epsilon=10^{-13}$ in the BSG algorithm, providing the relative energy error as small as $10^{-9}$. 

The second numerical model makes use of the fast indicator idea and relies in determining stability of the test orbits through their $\Ym$-signature, following arguments in Sect.~\ref{sec:Ym}.  We call the approach as the $\Ym$-model from hereafter.

The results of simulations, which are osculating orbital elements of stable systems, are projected onto the Cartesian coordinates $(x,y)$-plane, and also are shown in the semi-major axis--eccentricity $(a,e)$-plane. It helps to reveal and identify resonant structures in the phase space. 

We conducted two CPU-intensive tests, on up to 256 CPU cores, to validate the $\Ym$-model for debris disk through the direct numerical integrations. Computations are carried out for fixed masses of the bodies moving in the same plane. 

\subsection{The restricted four-planet system}

In the first test, including four planets, we conducted a number of \mercury{} runs with up to 4096 mass-less particles each. Our primary goal of this, and of further experiments, is to reconstruct the inner edge of the outer disk, hence we limit the semi-major axes to smaller range than predicted for the whole radius as large as $\sim 450$~au. Beyond the inner region of the disk, filled with strong MMRs, the population of asteroids might be determined with some quasi-analytic distribution \citep{Booth2016,Read2018}.  

In order to compare the results with findings in earlier papers, the initial conditions of the planets in this experiment are the same as in the best-fitting model IVa in \citep{Gozdziewski2014A}, also Tab.~\ref{tab:tab1}. We integrated the orbits of primaries and  mass-less particles for 34~Myr, which may be considered as the low limit of the system age. The simulation has been restricted to the inner part of the disk, i.e., the initial semi-major axes $a_0(t=0) \in[60,150]$~au. We sampled eccentricities $e_0(t=0) \in [0,0.33]$, as well as orbital phases randomly, with $\varpi_0,{\cal M}_0 \in [0^{\circ},360^{\circ})$.

\begin{figure*}
\centerline{ 
\vbox{
\hbox{
\includegraphics[width=0.33\textwidth]{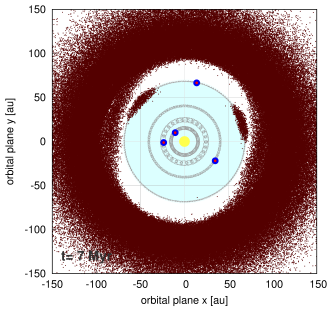}
\includegraphics[width=0.33\textwidth]{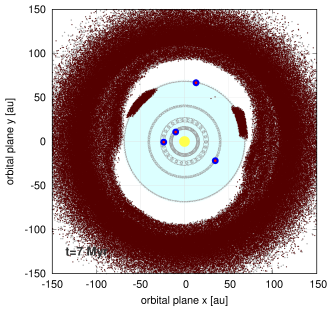}
\includegraphics[width=0.33\textwidth]{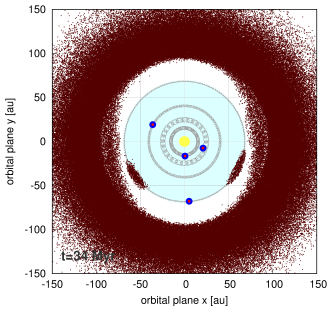}
}
\hbox{
\includegraphics[width=0.33\textwidth]{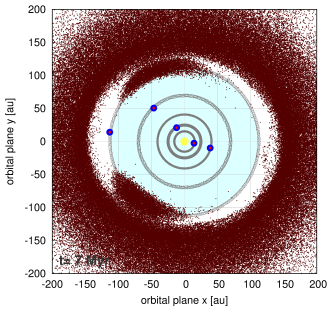}
\includegraphics[width=0.33\textwidth]{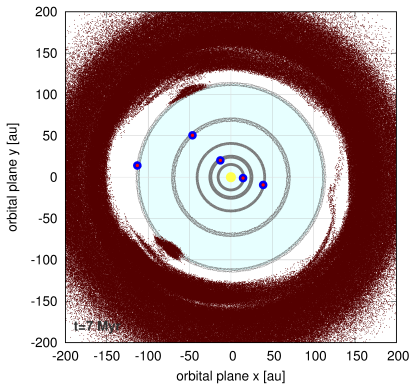}
\includegraphics[width=0.33\textwidth]{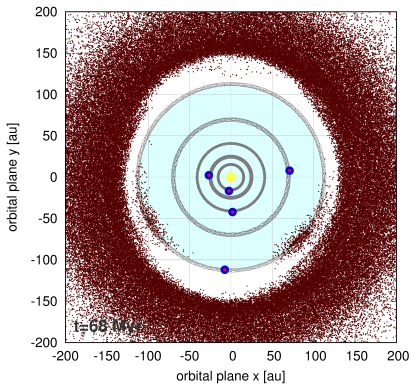}
}
\hbox{
\includegraphics[width=0.33\textwidth]{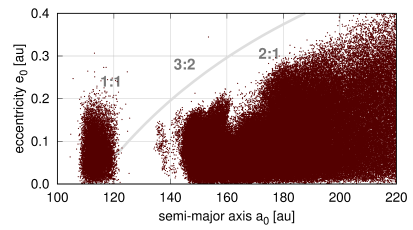}
\includegraphics[width=0.33\textwidth]{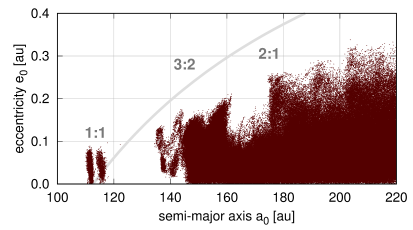}
\includegraphics[width=0.33\textwidth]{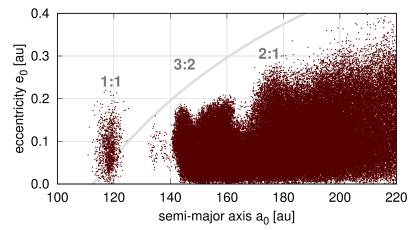}
}
}
}
\caption{
{\em Top row}:
the outer debris disk of the four-planet \hr8799{} system, in accord with model IVa in \citep{Gozdziewski2014A}. {\em The left-hand} and {\em the right-hand} panels show snapshots of astrocentric coordinates $(x,y)$ of mass-less asteroids at the orbital plane, at times $\simeq 7$~Myr and $\simeq 34$~Myr, respectively, derived with the direct $N$-body integrations conducted with the \mercury{} package (the hybrid scheme). Filled circles are for positions of the planets, and light-grey curves are orbital arcs sampled for the times-span $\simeq 34$~Myr.  The final snapshot involves $370,000$ mass-less particles which survived the integration. The middle panel shows astrocentric coordinates $(x,y)$ of $\simeq 10^6$ test particles with masses $10^{-15}~\mJ$ in $\Ym$-stable orbits. 
{\em Middle row}:
the outer debris disk of the five-planet \hr8799{} model~V$_{\rm A}$ (Tab.~\ref{tab:tab1}). {\em The left-hand} and {\em the right-hand} panels show snapshots of astrocentric coordinates $(x,y)$ of mass-less asteroids at the orbital plane, at the motion times $\simeq 7$~Myr and $\simeq 68$~Myr, respectively. Planet orbits are sampled and marked with grey dots for $\simeq 68$~Myr. The final snapshot involves $125,000$ mass-less particles which survived the integration. {\em The middle panel} is for astrocentric coordinates $(x,y)$ of $\simeq 1.5\times 10^6$ test particles with masses $10^{-15}~\mJ$ in $\Ym$-stable orbits.
{
Integrations done with the \mufarm{} package.
{\em Bottom row}:
Osculating, astrocentric Keplerian elements of the mass-less particles in the $(a_0,e_0)$-plane in the five planet model (the {\em middle row}). The light-grey curve marks the collision curve with planet \hr8799{}e. Approximate positions of low-order MMRs with this planet are labelled. 
}
}
\label{fig:fig7}
\end{figure*}

The total number of particles traced with the direct $N$-body integration was $\simeq 6\times 10^{5}$. Figure~\ref{fig:fig7} shows two snapshots of the simulation at $t=7$~Myr ({\em the top-left} panel), and at the end of the integration interval, $t=34$~Myr ({\em the top-right} panel). They represent instant coordinates of mass-less asteroids which have not been ejected beyond 1000~au. The final snapshot encompasses $3.7\times 10^{5}$ objects surviving the integration.

The \megno{} test is illustrated in {\em the top-middle} panel of Fig.~\ref{fig:fig7}. Almost $10^6$ particles with $\Y$ are marked at the integration time of $7$~Myr, which corresponds to $\sim$14,000 orbital periods of the outermost planet and roughly 8,000 revolutions at $\simeq 100$~au. Particles marked as $\Ym$-stable for that interval should persist for more than 10 times longer interval in Lagrange-stable orbits, roughly $60$--$70$~Myr. Indeed, the inner border of the debris disk derived with the direct \mercury{} integration looks very similar to the non-circular oval shape revealed by the $\Ym$-model. 

A different density of particles in the two snapshots may be explained by a sampling strategy. In regions, where the test orbits are strongly chaotic, like just beyond the orbit of \hr8799{}b, the motion is $\Ym$-unstable during short intervals $\simeq 0.1$~Myr, and the $\Ym$ integration may be stopped as soon as $\Y>5$, safely larger than $\Y\simeq2$ for stable systems. That made it possible to examine huge sets of $\sim 10^8$ initial conditions, orders of magnitude larger than they could be sampled with the direct $N$-body integrations. Such {strongly chaotic regions} are explored in a CPU-efficient way, and we argue that the inner, complex edge may be revealed more detailed with the $\Ym$-model.

We also note similarly extended Lagrange  islands $L_4$, $L_5$ of stable particles, indicating a sensitivity of the $\Ym$ indicator for unstable solutions.

The $\Ym$-model plot at the $(x,y)$-plane of Cartesian coordinates may be understood as a snapshot of {\em all possible}, stable (quasi-periodic, regular) orbits of the probe masses with various orbital phases and eccentricity for the same values of the semi-major axis. These orbits might be populated in a real system, but not necessarily they are. The actual population of asteroids may depend on the prior planetary system history, its migration, as well as locally variable density of asteroids.

\subsection{The restricted five-planet system}

We made also a second $\Ym$ calibration test with essentially the same settings, but for a preliminary five-planet model to the observations in \citep{Wertz2017} found with the \moa{}. Parameters of this solution are displayed in Tab.~\ref{tab:tab1} as model V$_{\rm A}$. In this model, a hypothetical, yet undetected $2.5\mJ$ planet at $a_{\rm f} \simeq 111.5$~au forms the 16e:8d:4c:2b:1f MMR chain with the observed planets. The test particles were integrated with the \mercury{} code for the interval of 68~Myr, even longer than before,  but for the same interval of $7$~Myr with the \mufarm{} MEGNO code. 

The results are shown in the bottom row of Fig.~\ref{fig:fig7}. We sampled asteroid orbits with $a_0(t=0) \in [60,220]$~au and $e_0 \in [0,0.4]$ implying the initial orbits up to the collision zone with orbit of planet~f. Also in this case, although the $\Ym$ interval is relatively short, the results of the direct $N$-body integrations and from the Monte Carlo $\Ym$ sampling closely overlap. Even subtle structures, such as a narrow arc made of particles opposite to the position of \hr8799{}b, are clearly present in {\em the two last} panels. Also the overall, egg-like shape of the inner edge looks the same. Initially massive $L_4$ and $L_5$ Lagrangian clumps of asteroids in the {\rm bottom-left} panel are finally reduced to smaller, dispersed islands with similarly wide centers, as seen in the {\em bottom-right} panel. These islands appear as much more compact structures in the $\Ym$-model plot ({\em bottom-middle} panel). This implies that particles initially forming a kind of echo around $L_4$ and $L_5$ co-rotation centers in the direct integration plot ({\em bottom-left} panel) are secularly unstable. They are eventually removed from the system.

{
We projected the Keplerian elements of the test particles at the $(a_0,e_0)$-plane, as shown in {\em the bottom row} of Fig.~\ref{fig:fig7}. The distribution of the elements derived with short-term $\Ym$ integrations ({\em the bottom-middle} panel) closely matches with data after $68$~Myr found directly  with the \mercury{} code. We note a two-modal structure of the 1:1 MMR ({\em the bottom-midle} $\Ym$ panel), which has been detected with a dense sampling of the particles. The distribution of elements appears diffuse due to their representation in the common, astrocentric Keplerian frame \citep{Lee2001}. It hinders the true, resonant structure of the disk, and we address this issue further in Sect.~\ref{sec:planetV} and \ref{sec:outer}.
}

{One should be aware that the $\Ym$-model regards the short-term resonant dynamics \citep[e.g.,][]{Laskar2001}, and the relatively short integration times may not be sufficient to resolve long-term secular resonances present in all systems with more than two planets \citep[e.g.,][]{Morbidelli2001}. However, since the obtained $\Ym$ distribution of elements overlap with the results of the direct $N$-body integrations, the dynamical effects of the secular resonances are either non-detectable during $\sim 70$~Myr or overlap with the short-term MMR dynamics.}

\section{The inner disk dynamical structure}
\label{sec:inner}

In order to even better validate the $\Ym$-model, we compared its outcomes with the results in a recent work by \cite{Contro2016}, regarding the inner debris disk in the \hr8799{} system \citep{Su2009,Reidemeister2009,Hinkley2011, Matthews2014}. Its structure is not yet fully resolved. 

We conducted experiments for varied mass of test particles: small, essentially mass-less asteroids of $m_0=10^{-15} \mJ$, a super-Earth with $m_{\rm f}=10$~Earth masses, and $m_{\rm f}=1\mJ$ Jovian planet, respectively. We investigated a zone beyond $4$~au from the star up to 2~au beyond the inner orbit of \hr8799{}e ($\simeq 17$~au), and orbits with $e_{0,\rm f} \in [0,0.4]$, roughly within the collision zone with the orbit of planet~e. As before, we aim to resolve the outer edge of the disk which is carved by the closest planet and by the whole system indirectly, through the coupled resonant motion. The initial conditions for the primaries are the same as in our new astrometric model IV$_{\rm K}$ (Tab.~\ref{tab:tab1}).

\begin{figure*}
\centerline{ 
\vbox{
\hbox{
\includegraphics[width=0.33\textwidth]{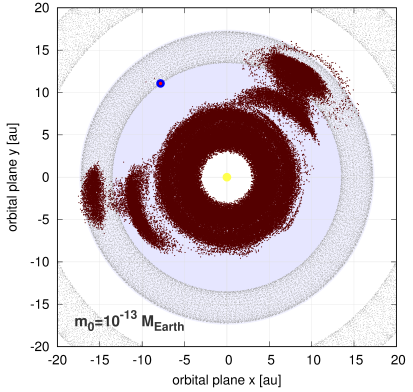}
\includegraphics[width=0.33\textwidth]{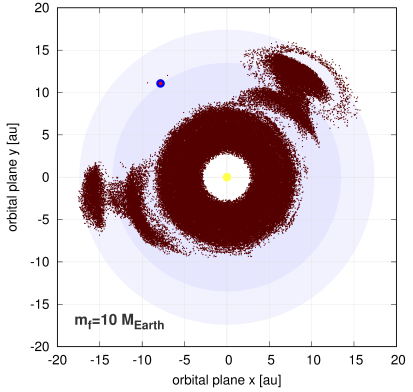}
\includegraphics[width=0.33\textwidth]{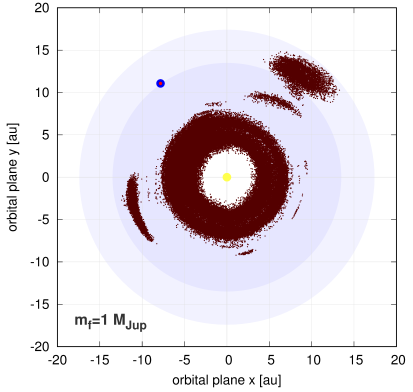}
}
\hbox{
\includegraphics[width=0.33\textwidth]{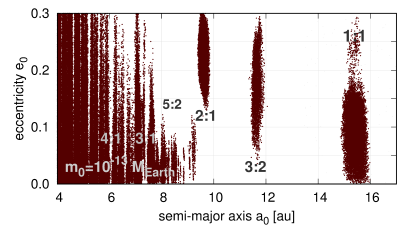}
\includegraphics[width=0.33\textwidth]{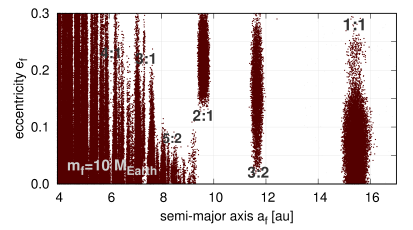}
\includegraphics[width=0.33\textwidth]{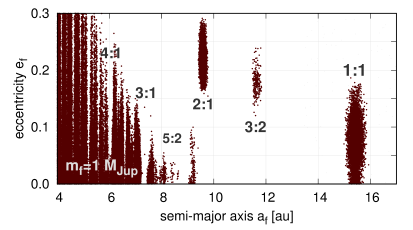}
}
\hbox{
\includegraphics[width=0.33\textwidth]{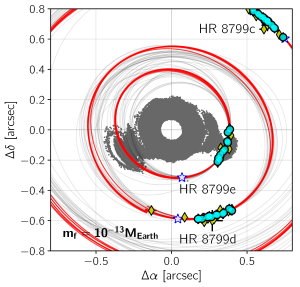}
\includegraphics[width=0.33\textwidth]{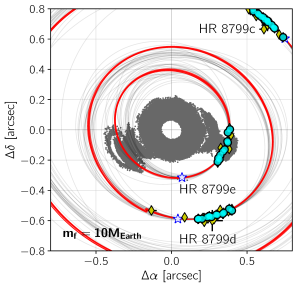}
\includegraphics[width=0.33\textwidth]{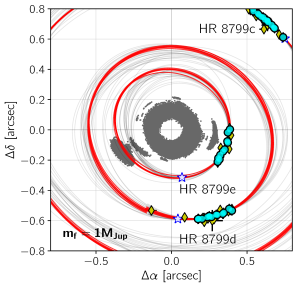}
}
}
}
\caption{
The inner debris disk ({\em the left} column) and possible orbits of yet undetected innermost planet {\hr8799{}f} ({\em the middle} and {\em right} columns) in model IV$_{\rm K}$ at the initial osculating epoch \epk{}.  Panels in {\em the top} row  are for instant astrocentric $(x,y)$ coordinates for different masses of test bodies in stable orbits.  {\em The left} column is for $10^{-15}\mJ$, {\em the middle} column is for $10$M$_{\rm Earth}$, and {\em the right} column is for $1$M$_{\rm Jup}$, respectively.  A filled circle marks the initial position of planet \hr8799{}e.  Light-grey dots in the top-left panel illustrate  orbits of two inner planets sampled for the integration time of 3~Myr. For a reference, light-blue discs in two next panels mark the nominal periastron and apoastron distance of \hr8799{}e. 
Panels in {\em the middle} row are for the $(a_0,e_0)$- or $(a_{\rm f},e_{\rm f})$-planes of the initial Keplerian elements with labels identifying low-order MMRs with the innermost planet \hr8799e{}.
Panels in {\em the bottom} row are for the $(x,y)$ coordinates rotated to the sky plane, in which the $y$-axis corresponds to N, and the $x$-axis corresponds to E (note the numerical values of $\Delta\alpha$ are sign-opposite to with respect to the formal left-hand direction of the right ascension $\alpha$). Astrometric observations in \citep{Konopacky2016} are marked with filled circles, and other measurements listed in \citep{Wertz2017} are marked with diamonds. Several models selected from the MCMC sampling are marked with red (stable, $\Chi\simeq 1$) and grey curves (unstable, $\Chi<0.74$), respectively. For a reference, positions of planets \hr8799{}c,d,e at the initial epoch of \epk{} are marked with a~star symbol. Note a variability of unstable models, in spite of relatively small $\Chi$.
Snapshots in subsequent panels illustrate $\sim 250,000$, $\sim 240,000$ and $\sim 220,000$ $\Ym$-stable orbits, respectively, for 3~Myr ($\simeq 6\times 10^4$ and $\simeq 3\times 10^4$ revolutions of the two closest perturbers \hr8799{}e,d).
}
\label{fig:fig8}
\end{figure*}

The results are illustrated in Fig.~\ref{fig:fig8}. In all panels, the innermost planet \hr8799{}e is marked with a filled circle. 

The {\em left-hand} column of panels is for the restricted problem. The orbits of planets \hr8799{}e,d are sampled and marked with gray dots for the integration interval of 3~Myr. That corresponds to 40,000 revolutions of the innermost perturber \hr8799{}e and for more than 4,000 revolutions of the outermost planet~b. The $(x,y)$ plots, representing stable orbit s at the initial osculating epoch, are accompanied by the final distribution of the astrocentric Keplerian elements in the $(a_0,e_0)$-plane for mass-less objects and in the $(a_{\rm f},e_{\rm f})$-plane for non-zero mass planets (the two remaining columns).

Although a mass of the fifth body is varied, {the inner disk reveals similar features determined by low-order MMRs} with the innermost planet~e. By including mass-less particles with moderate eccentricities $e_0 \in [0,0.4]$ in the initial distribution, we obtain {highly asymmetric shape of the outer parts of the disk}, with large Lagrangian $L_4,L_5$ clumps accompanied by complex structures of the 3:2~MMR with the innermost planet~e. We note that the overall border of the disk edge is different from that found by \cite{Contro2016}, who also mapped the phase space with the $\Ym$ indicator. By fixing initial phases of the asteroids in 2-dim scans one obtains non-exhaustive representation of stable regions. For instance, the 3:2 MMR is missing at $(a_0,e_0)$ scans shown in \citep{Contro2016}.

The $\Ym$-model is also useful to ``predict'' positions of a hypothetical innermost, yet undetected fifth planet~``f'' in the system. Such a body has been considered as an explanation of the observed Spectral Energy Distribution (SED) in \citep{Su2009,Hinkley2011}. In \citep{Gozdziewski2014A}, we simulated such a body with the \moa{} algorithm, and we found a few possible locations of the missing planet, associated with low order 2:1 and 3:1~MMR with \hr8799{}e. 

However, the gravitational influence of such a hypothetical planet on the inner companion \hr8799{}e could be hardly detected through deviations of astrometric measurements, given short orbital arcs and relatively large uncertainties.  We can simulate potential locations of the hypothetical planet more efficiently with the $\Ym$-model, attributing non-zero mass to the ``asteroids''. We conducted such experiments for $m_{\rm f}=10$~Earth masses and for $m_{\rm f}=1 \mJ$ ({\em the middle } and {\em the right-hand} columns in Fig.~\ref{fig:fig8}, respectively). The most noticeable phase-space structures are preserved and look similar to the restricted case. The additional, putative planet should be involved in 2-body MMRs with the inner observed planet \hr8799{}e at orbits down to $a_{\rm f} \simeq 6$~au. It also means that the four-planet model is robust against perturbations. These predictions overlap with our earlier simulations in \citep{Gozdziewski2014A}. 

We note that attempts to detect or dismiss the hypothetical inner fifth planet failed so far. This planet is below detection limits due to small mass or too close proximity to the star and insufficient contrast \citep{Skemer2012,Matthews2014,Maire2015,Zurlo2016}.

Bottom panels in Fig.~\ref{fig:fig8} are for the sky-projected snapshots of stable orbits at the initial epoch of \epk{} (for a reference, marked with a star), overplotted with several model orbits from the MCMC sampling and all measurements in \citep{Wertz2017}. These plots might be useful in interpreting the direct imaging observations. Down to $\simeq 7$--8~au ($\simeq 0''.2$), the ``missing'' planet may be found in discrete, isolated islands associated with 3:2, 2:1, 5:2 and 3:1 MMRs with planet~e. Below that limit, it could persist essentially everywhere in a wide ring around the star.

%
\section{The outermost planet V hypothesis}
%
\label{sec:planetV}

\begin{figure*}
\centerline{ 
\vbox{
\hbox{
\includegraphics[width=0.33\textwidth]{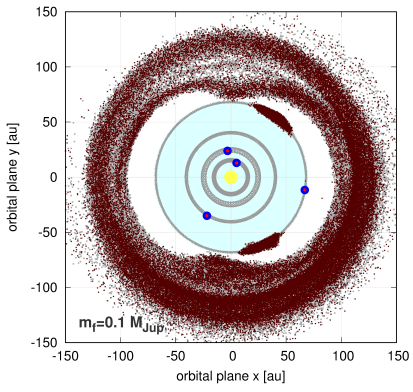}
\includegraphics[width=0.33\textwidth]{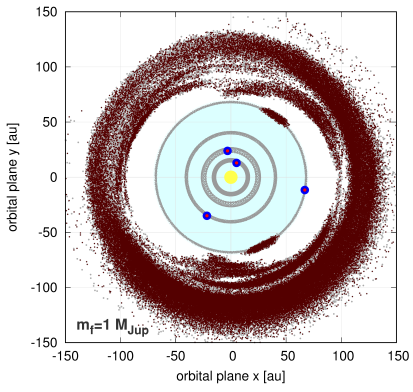}
\includegraphics[width=0.33\textwidth]{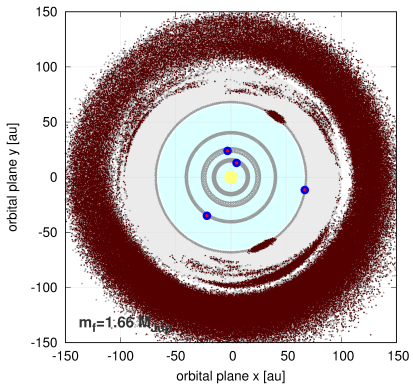}
}
\hbox{
\hspace*{2.mm}\includegraphics[width=0.33\textwidth]{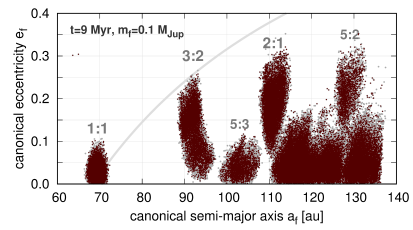}
\hspace*{0.mm}\includegraphics[width=0.33\textwidth]{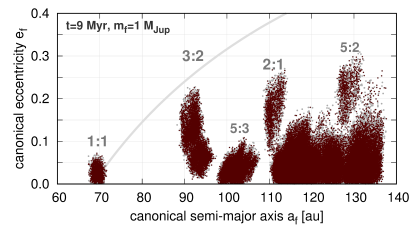}
\includegraphics[width=0.33\textwidth]{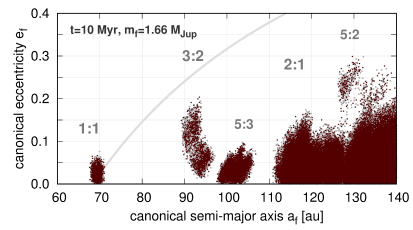}
}
\hbox{
\includegraphics[width=0.33\textwidth]{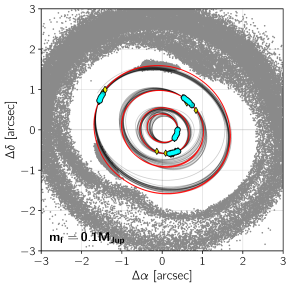}
\includegraphics[width=0.33\textwidth]{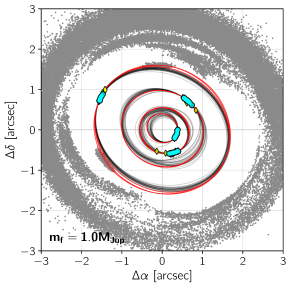}
\includegraphics[width=0.33\textwidth]{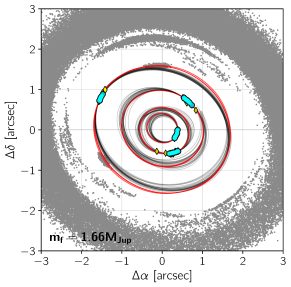}
}
}
}
\caption{
The $\Ym$-model simulations of localizations of a putative planet \hr8799{}f in the outer region of the system. Subsequent panels show astrocentric coordinates $(x,y)$ of stable orbits in the orbital plane ({\em top row}), a distribution of the Poincar\'e canonical orbital  elements $(a_{\rm f},e_{\rm f})$ ({\em the middle row}) and a projection of the instant coordinates at the sky plane, relative to the observed system (filled circles) and several four-planet synthetic orbits from the MCMC sampling, marked with red and grey curves for stable and unstable models, respectively. The $y$-axis corresponds to N, and the $x$-axis corresponds to E (note that the numerical values of $\Delta\alpha$ are sign-opposite with respect the formal left-hand direction of the right ascension $\alpha$).
Columns are for different masses of the putative planet of $0.1 \mJ$, $1 \mJ$ and $1.66 \mJ$, respectively.  Light-grey curves in the middle panels indicate the collision curve for planet \hr8799{}e and test objects. Integrations are done with the \mufarm{} package with the \megno{} indicator for 9~and 10~Myr. Grey rings in {\em the top panels} illustrate sampled orbits of the observed planets for the integration time.  The snapshots show up to $2\times 10^5$ objects in $\Ym$-stable orbits. The reference epoch \epk{} is the date of the first observation in \citep{Konopacky2016}.
}
\label{fig:fig9}
\end{figure*}

The inner edge of the outer cold debris disk has been detected at $145$~au, hence much farther than $\simeq 90$~au that could be explained by the gravitational pull of planet \hr8799{}b \citep{Booth2016,Read2018}.  With the same framework and orbital model IV$_{\rm K}$ of the primaries, as in the previous section regarding the inner disk structure, we aim to investigate locations of a hypothetical fifth planet \hr8799{}f exterior to \hr8799{}b. We computed the $\Ym$-model for a few choices of the mass $m_{\rm f}$ of the putative planet. The results are illustrated in Fig.~\ref{fig:fig9}, for $m_{\rm f}=0.1 \mJ$, $m_{\rm f}=1 \mJ$, and $m_{\rm f}=1.66 \mJ$, respectively, as well as we made a similar experiment for $m_{\rm f}=2.5 \mJ$ (not shown). 

The top row in Fig.~\ref{fig:fig9} illustrates possible, stable positions of \hr8799f{} in the orbital plane at the initial epoch of \epk{}. Again, it should be understood as a representation of the initial conditions implying stable evolution of the system, rather than a physical disk. The distribution of orbits is very similar to that one derived for mass-less or low-mass asteroids shown in Fig.~\ref{fig:fig7}. In order to understand the dynamical structure of this quasi-disk, we also plotted the same orbits projected onto the $(a_{\rm f},e_{\rm f})$-plane ({\em the middle-row} panels), where $a_{\rm f}$ and $e_{\rm f}$ are the canonical, Poincar\'e elements \citep{Morbidelli2001,Gozdziewski2008}. This parametrization of the orbital elements is critical to ``disentangle'' the structure of MMRs. Otherwise, a classification of orbits is obscure with the common, Keplerian astrocentric elements due to significant variation of the osculating semi-major axes of the outer planets \citep[for instance,][]{Lee2001}, see also {\em the bottom row} in Fig.~\ref{fig:fig6}. 

As can be deduced from Fig.~\ref{fig:fig9}, a Neptune-mass planet could belong to a stable system when involved in a low order 2-body resonance with \hr8799{}b, like the 1:1, 3:2, 5:3, 2:1 and 5:2 MMRs. In that instance, the eccentricity of the fifth planet could be as large as $e_{\rm f}\simeq 0.3$. It might be found at essentially  any location at the sky (see {\em the bottom-row} panels), beyond the angular distance roughly equivalent to the semi-major axis of $a_{\rm f} \simeq 90$~au. Also two Trojan, 1:1~MMR locations are possible. A similar, extended yet more sparse and clear pattern of the MMRs is present for a larger mass of $1 \mJ$ ({\em the middle} column), as well as for $1.66 \mJ$ ({\em the right} column), and $2.5 \mJ$ (not shown).

%
\section{The outer debris disk structure}
%
\label{sec:outer}

Finally, we conducted $\Ym$-model simulations for a few copies of the \hr8799{} best-fitting solution, model IV$_{\rm K}$ in Tab.~\ref{tab:tab1}, involving the fifth planet with different masses and the initial semi-major axis. As shown in the previous section, such a planet is weakly constrained by the present astrometry. However, the recent observations of the outer disk imply conjugated constrains for its structure as well as for possible orbit and mass of this putative object. 

We aim to simulate the outer disk composed of probe masses $m_0$ set to $10^{-15}\mJ$ in further experiments.  After selecting a mass and orbital elements of the fifth, additional planet \hr8799{}f from simulations described in Sect.~\ref{sec:planetV} and illustrated in Fig.~\ref{fig:fig9}, we verified $\Ym$ scans in the $(a_{\rm f},e_{\rm f}$)-plane. We checked whether its stability zone is sufficiently wide, to avoid biases due to a proximity to unstable resonances. When necessary, the semi-major axis $a_{\rm f}$ has been slightly modified to separate it safely from all nearby unstable MMRs. The orbital elements of the primaries selected for computations are gathered and labeled in Tab.~\ref{tab:tab1}, as \hr8799{}fA to \hr8799{}fE, complementing model IV$_{\rm K}$. We tested orbits with the initial $a_{0,{\rm k}} \in [60,240]$~au and eccentricities $e_{0,{\rm k}} =[0,\sim 0.5)$ below the collision curve with the outer planet. (For the mass-less particles, we distinguish the Poincar\'e canonical elements $(a_{\rm p},e_{\rm p})$ from $(a_0,e_0) \equiv (a_{\rm k},e_{\rm k})$ denoting common, Keplerian astrocentric elements). 

\begin{figure*}
\centerline{ 
\vbox{
\hbox{
\includegraphics[width=0.33\textwidth]{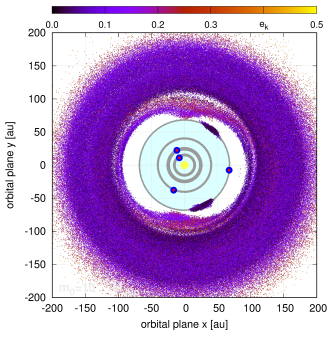}
\includegraphics[width=0.33\textwidth]{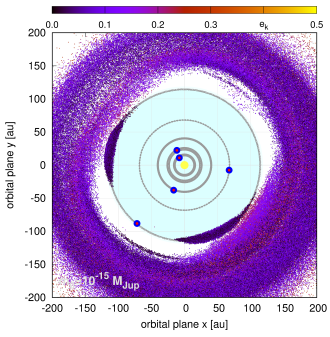}
\includegraphics[width=0.33\textwidth]{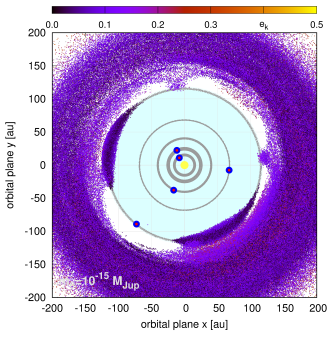}
}
\hbox{
\includegraphics[width=0.33\textwidth]{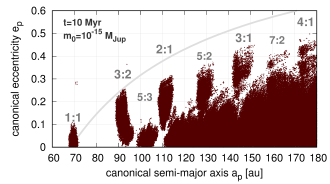}
\includegraphics[width=0.33\textwidth]{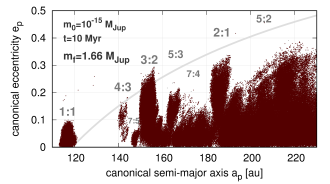}
\includegraphics[width=0.33\textwidth]{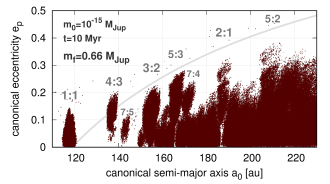}
}
}
}
\caption{
The inner part of the outer debris disk of the four-planet ({\em the left} column) and five-planet systems (remaining two columns). Orbital elements of the fifth planet, complementing model IV$_{\rm K}$, are listed as \hr8799{}fA and \hr8799{}fB, respectively, in Tab.~\ref{tab:tab1}. Subsequent panels in {{\em the top}} row show astrocentric coordinates $(x,y)$ of asteroids in $\Ym$-stable orbits, for different masses of a hypothetical companion \hr8799{}f,  at the initial epoch of \epk{}. Panels in {{\em the bottom}} row are for the canonical, Poincar\'e  elements $(a_{\rm p},e_{\rm p})$ of these stable solutions at the end of the integration interval of 10~Myr. Grey lines are for the collision curve of orbits with planet \hr8799{}f. The low-order MMRs with planet \hr8799{}f are labeled in bottom panels. Osculating, initial orbital eccentricities $e_{\rm k}$ at the initial epoch are color-coded in {\em the top} row. Snapshots in subsequent panels illustrate $\sim 920,000$, $\sim 580,000$ and $\sim 910,000$ $\Ym$-stable orbits, respectively.
}
\label{fig:fig10}
\end{figure*}

The first set of simulations is illustrated in Fig.~\ref{fig:fig10}.  Here, we essentially repeated the $\Ym$ calibrating experiment in Sect.~\ref{sec:calibration}, yet to determine the edge of the outer disk with the updated model IV$_{\rm K}$. The ($x,y$) snapshot of stable orbits at the initial osculating epoch of \epk{} is now complemented with the $(a_{\rm p},e_{\rm p})$-plane for the canonical, Poincar\'e elements of the orbits of test particles at the end of the integration interval of $10$~Myr, extended to more than $10^4$ orbits at $\sim 100$~au. The integration time implies that orbits characterized as $\Y$-stable should be Lagrange-stable for $\simeq 100$~Myr. The $(x,y)$ orbits are coded in a colour-scale representing their initial, Keplerian astrocentric eccentricity $e_{\rm k}$. This provides an additional information on the distribution and geometry of stable orbits at the initial epoch of \epk{}.

The ($x,y)$-orbital plane shown in the top-left panel in Fig.~\ref{fig:fig10} reveals similar features to those seen  in Fig.~\ref{fig:fig7} ({\em the top} row). A shape of the inner edge of the disk, resembling a fat peanut, is strongly distorted by orbits of asteroids in the 3:2  and 5:3~MMRs with \hr8799{}b. They may be easily identified in the $(a_{\rm p},e_{\rm p})$-plot. There are also present large islands of stable particles co-rotating with the $L_4,L_5$ -Lagrangian points of planet~b. They extend for wide arcs as long as $\simeq 40$~au. The short-term dynamics in the 100~Myr time-scale is apparently governed by the major gravitational pull of \hr8799{}b, since stable orbits are possible essentially only below the collision curve $e_{\rm p}(a_{\rm p})$, marked in the {\em bottom} row panels in grey. This curve is determined from $a_{\rm b} (1+e_{\rm b}) \simeq a_{\rm p} (1-e_{\rm p})$. Some proportion of the asteroids might move on moderately eccentric orbits up to $e_{\rm p}\simeq 0.4$ when trapped in low-order MMRs.

Two next columns in Fig.~\ref{fig:fig10} are for five-planet systems with   planet \hr8799{}fA of mass $m_{\rm f}=1.66 \mJ$ and  planet \hr8799{}fB of mass $m_{\rm f}=0.66 \mJ$, respectively (see model IV$_{\rm K}$ in
Tab.~\ref{tab:tab1}). The semi-major axis $a_{\rm f}\sim${116}~au, forming the 16e:8d:4c:2b:1f MMR chain, is
{almost the same} in both experiments. These configurations address the inner edge at $\simeq 145$~au, as detected in \citep{Booth2016}. Indeed, for $m_{\rm f}=1.66 \mJ$, stable orbits exhibit semi-major axes $a_{\rm p}>140$~au, beyond Lagrangian $L_4,L_5$ zones, with two strongly resonant regions of the 3:2 and 2:1 MMRs with planet \hr8799{}f. A similarity of the $(x,y)$ distribution with the results in {\em the bottom} panels of Fig.~\ref{fig:fig7} is striking. The inner edge would be strongly distorted by moderately eccentric orbits $e_{\rm p} \simeq 0.2$ associated with the 3:2~MMR. A ring of eccentric 2:1 MMR orbits with $a_{\rm p} \simeq 150$~au is marked with reddish points in {\em top-middle} panel. The Lagrangian zones are much wider than in the four planet configuration (panels in {\em the left} column of Fig.~\ref{fig:fig10}) and extend along arcs of $\sim 100$~au, contributing to even more non-circular and asymmetric inner edge of the disk.

Even more complex inner shape appears for smaller mass $m_{\rm f}=0.66 \mJ$ of the hypothetical planet \hr8799{}f at $a_{\rm f} \simeq 116$~au ({\em the right column}). While the inner edge in the $(a_{\rm p},e_{\rm p})$-plane is similar to the $m=1 \mJ$ case, now the 4:3~MMR is populated with more particles. Also the Lagrange zones are more extended, and moderately eccentric orbits in the 4:3~MMR form two clumps opposite to the planet, at distances equal to its mean heliocentric semi-major axis $a_{\rm f}$. Other features of the asteroids distribution are similar to the previous case.

\begin{figure*}
\centerline{ 
\vbox{
\hbox{
\includegraphics[width=0.33\textwidth]{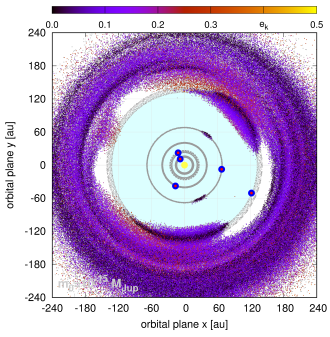}
\includegraphics[width=0.33\textwidth]{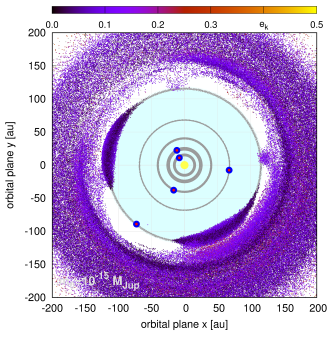}
\includegraphics[width=0.33\textwidth]{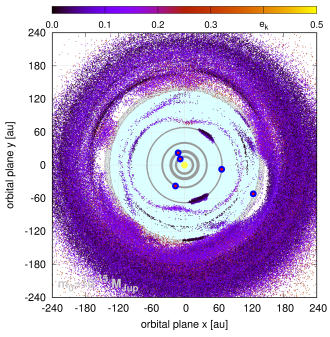}
}
\hbox{
\includegraphics[width=0.33\textwidth]{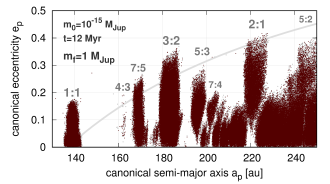}
\includegraphics[width=0.33\textwidth]{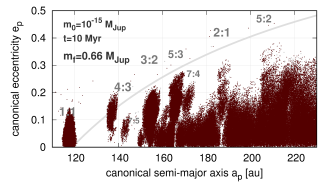}
\includegraphics[width=0.33\textwidth]{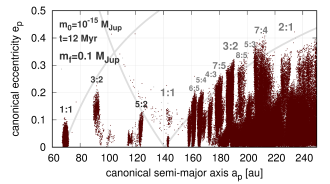}
}
}
}
\caption{
The inner part of the outer debris disk of the five-planet systems and different masses $m_{\rm f}=1 \mJ$, $m_{\rm f}=0.33 \mJ$ and $m_{\rm
f}=0.1 \mJ$, respectively,  of a hypothetical, fifth planet beyond the orbit of  \hr8799{}b. Orbital elements of the fifth planet, complementing model IV$_{\rm K}$, are listed as \hr8799{}fC, \hr8799{}fD and \hr8799{}fE, respectively, in Tab.~\ref{tab:tab1}. A mass of this additional planet is labeled at the bottom panels. Subsequent panels in {{\em the top}} row show astrocentric, Cartesian coordinates $(x,y)$ of asteroids at $\Ym$-stable orbits, for different masses $m_{\rm f}$ of a hypothetical companion \hr8799{}f labeled in subsequent bottom panels, at the initial epoch of \epk{}. Panels in {\em the bottom} row are for the canonical, Poincar\'e elements of these stable solutions at the end of the integration intervals of 10~Myr and 12~Myr. Grey lines are for collision curves of orbits with planet \hr8799{}f, and also with planet  \hr8799{}b ({\em the right} column). Low-order MMRs with planet \hr8799{}f, and with planet \hr8799{}b are labeled. Osculating astrocentric eccentricities $e_{\rm k}$ at the initial epoch are color-coded in panels at {\em the top} row. Snapshots in subsequent panels illustrate $\sim 980,000$, $\sim 635,000$ and $\sim 500,000$ $\Ym$-stable orbits, respectively.
}
\label{fig:fig11}
\end{figure*}

We also considered an orbital setup with the outer planet~f beyond $\simeq 130$~au, as predicted by the disk models in \citep{Booth2016,Read2018}. Due to the larger semi-major axis of the perturber, the $\Ym$ integration time has been increased to 12~Myr ($\sim 10^4$ orbits of the outermost perturber at $\sim 130$~au). We investigated three cases with $m_{\rm f}=1 \mJ$,  $m_{\rm f}=0.33 \mJ$, and $m_{\rm f}=0.1 \mJ$, respectively, illustrated in columns of Fig.~\ref{fig:fig11}. While the overall shape of the disk is roughly similar to the previous models, new features appear. The inner edge becomes more and more distorted from a circle for decreasing masses $m_{\rm f}$. Simultaneously, clumps of stable orbits associated with the Lagrangian equilibria develop to a state in which they are merged with the inner edge at the mean distance equal to the semi-major axis $a_{\rm f}$. Also, the low order MMRs zones are even better isolated, hence the inner edge up to $\simeq 220$~au is strictly resonant.

The most complex image of the inner border of the disk has been found for the smallest mass of $m_{\rm f}=0.1 \mJ$ placed at $a_{\rm f} \simeq 138$~au. This particular configuration follows the best-fitting solution explaining the recent ALMA observations in \citep{Read2018}. The resonant structure associated with this planet is also very sharp, yet some particles may survive in stable, resonant orbits between planets \hr8799{}b and \hr8799{}f. We did not find such orbits for $m_{\rm f}=0.33 \mJ$, although planet~f is shifted by only $\simeq 5$~au. The inner resonant orbits are associated with 1:1, 3:2 and 5:2~MMRs with planet \hr8799{}b. In both zones, the eccentricities might be excited beyond collision curves with planets \hr8799{}b and \hr8799{}f.

%
\section{The debris disc under migration}
%
\label{sec:migration}

The resonant nature of the four known planet configuration might result from the migration due to the planet-disc interactions. Including this process in the debris disc formation scenario may potentially shift the border with respect to the results of the previously studied models in which all the planets are statically placed at their current orbits together with a disc of the mass-less particles. We therefore aim to investigate whether migration of the planets might place the inner edge of the outer debris belt at $\sim 145$~au. Then the requirement for an additional planet outside the orbit of planet~b could be released. While the analysis is preliminary, it might offer an alternative scenario for explaining the ALMA observations.

We assume that the planets are initially located outside their current positions and they undergo the inward convergent migration that is supposed to form a chain of 2:1~MMRs. We use the same model of the planet-disc interactions, as used before and described by Eq.~\ref{eq:migration}, {yet with $\kappa$ the same for all planets}.

Although the mass-less particles are too small to undergo the type~I migration, they could, in principle, ``feel''' the gas drag \citep{Adachi1976}. The timescales of the orbit decay that results from the drag depends strongly on the particles masses as well as on the gas densities (the volume densities, not the surface densities as for the type~I and type~II migration). Assuming that the surface gas density $\propto r^{-3/2}$ and the total gas mass within the $(10, 100)\,\au$ range equals $1\,\mJ$ when we start the simulation, the volume gas density would be in a range of $10^{-14}$ and $10^{-15}$g/cm$^3$, depending on the assumed aspect ratio between $0.01$ and $0.1$. For particles larger than a meter the characteristic timescale of the orbit decay induced by the gas drag would be greater than the age of the system, and greater that the migration timescales assumed for the planets, that are $\sim 10$Myr. Therefore, we neglect the migration of the mass-less particles in this model.

We performed a series of simulations for different initial orbits and the migration parameters.  We assume that the resulting system has to be resonant, of a similar size as the observed configuration and the inner edge of the mass-less particles disc has to place in the range of $(145 \pm 12)\,\au$ as determined by \cite{Booth2016}. These constrains do not permit for very wide initial orbits, since the resulting inner edge would be placed outside the observed border. On the other hand, all the pairs of planets need to be located outside 2:1~MMRs, as the resonances are formed in the process of the convergent migration. Moreover, the innermost planet should start the migration beyond its current position at $\sim 15.5\,\au$. 

\begin{figure*}
\centerline{
\vbox{
\hbox{
\includegraphics[width=0.38348\textwidth]{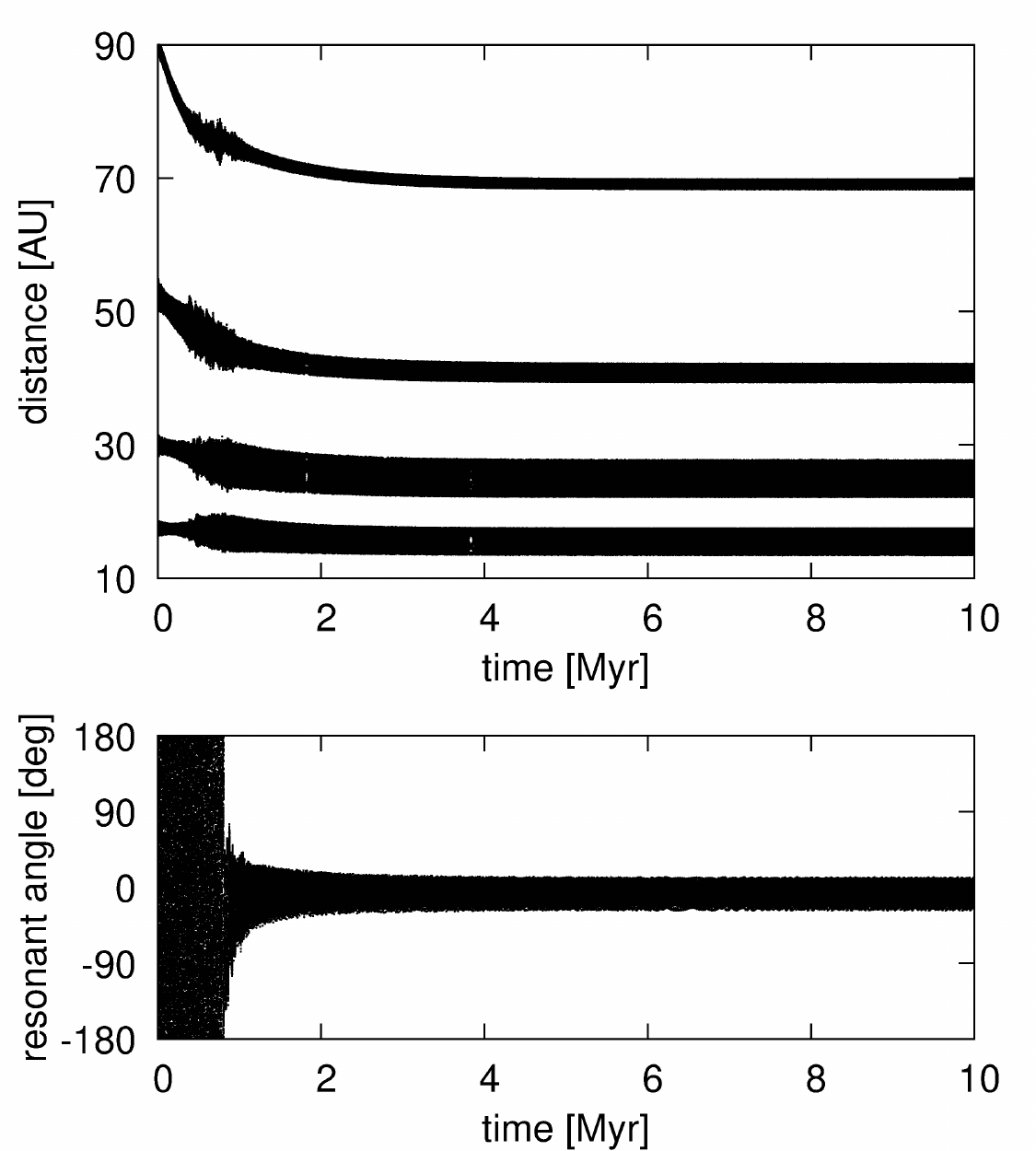}
\includegraphics[width=0.29826\textwidth]{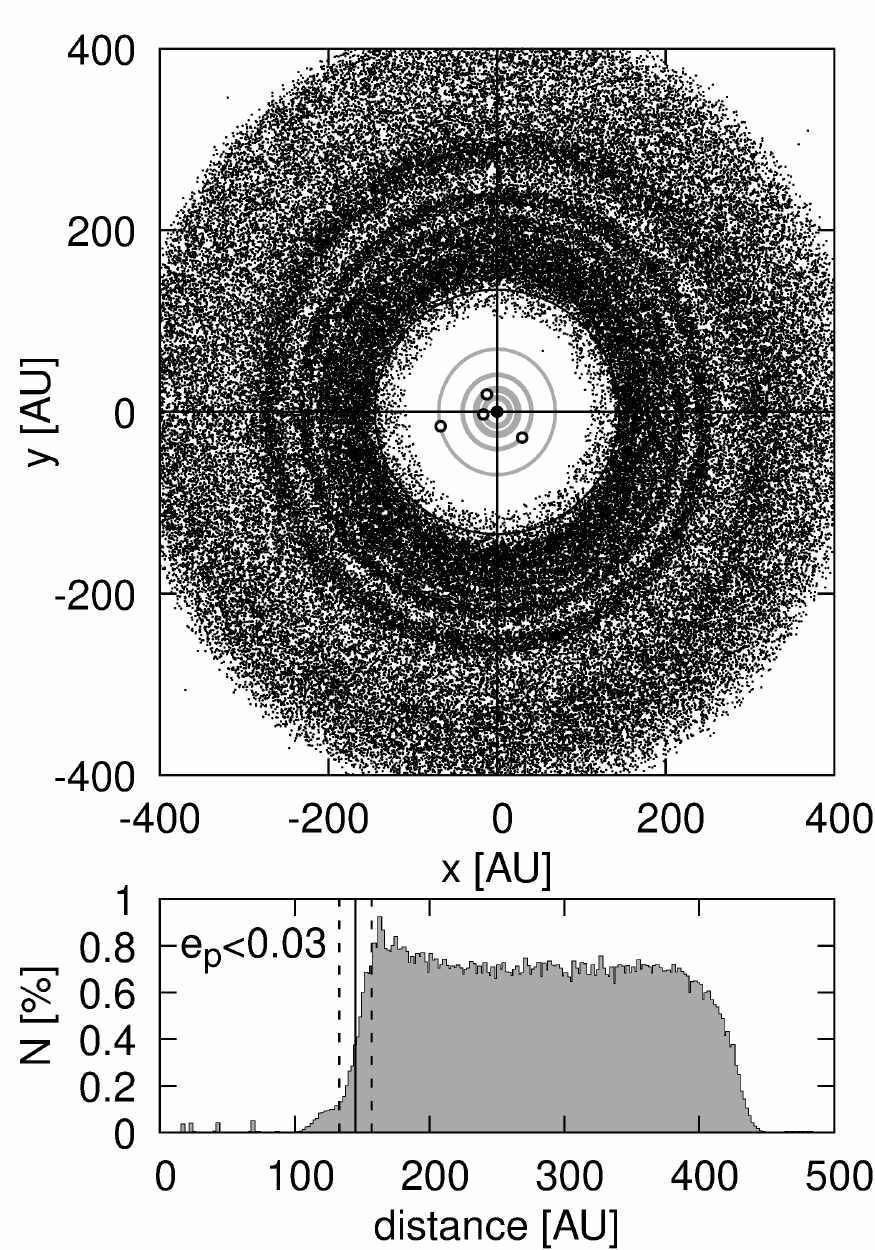}
\includegraphics[width=0.29826\textwidth]{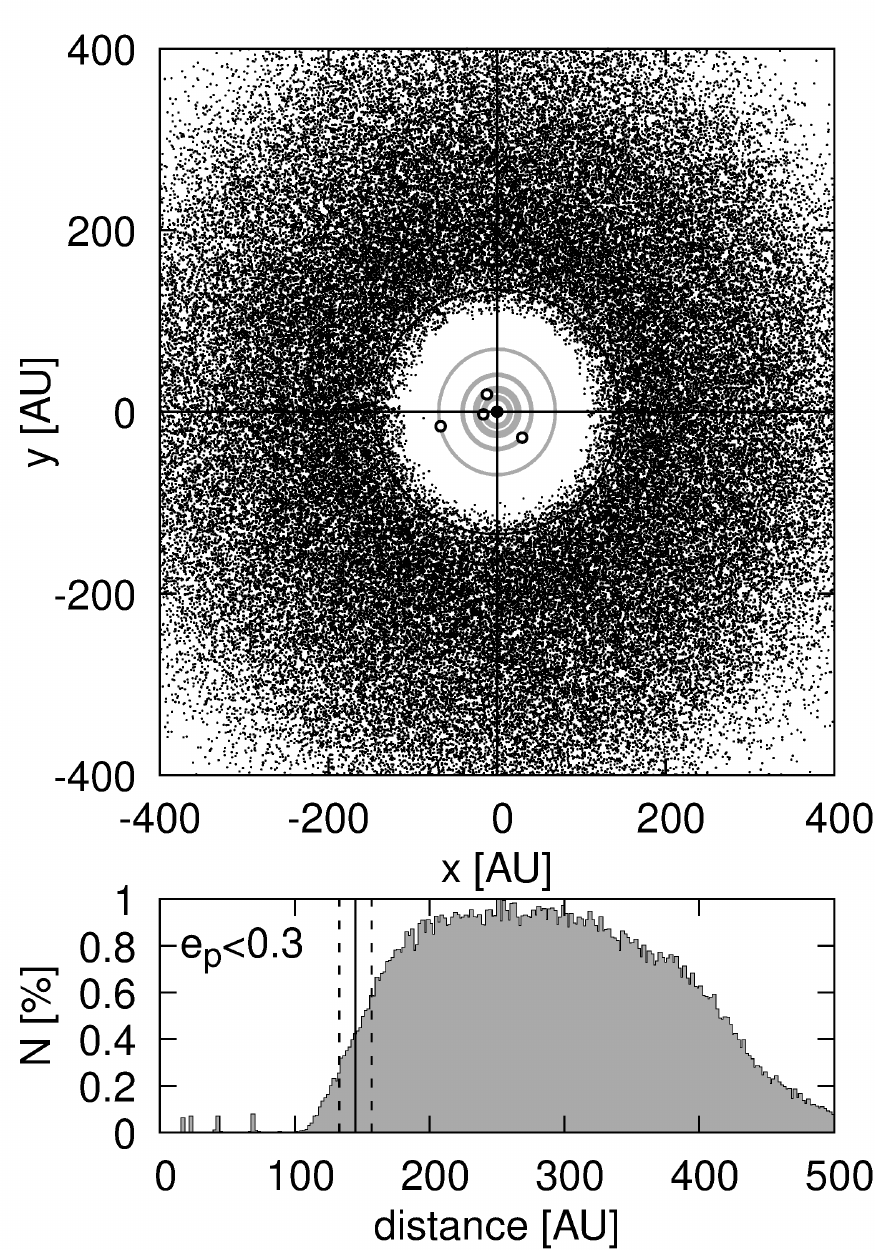}
}
}
}
\caption{{\em The left-hand column}: Evolution of an example initial configuration that ends up as a system resembling the configuration of the \hr8799{}. Initial semi-major axes are $a_1 = 17.5\,\au, a_2 = 30\,\au, a_3 = 52\,\au$~and $a_4 = 90\,\au$. All eccentricities and arguments of pericenters are initially $0$. Mean anomalies are chosen to be $0, 90, 180$~and $170$~degrees, from the innermost to the outermost planets respectively. The migration parameters are $T = 1~$Myr, $\kappa = 200$ and respective timescale of migration $\tau_1 = 28~$Myr, $\tau_2 = 18~$Myr, $\tau_3 = 12~$Myr, $\tau_4 = 9~$Myr. The masses are the same as in solution IVa in \citep[][see also Tab.~\ref{tab:tab1}]{Gozdziewski2014A}, i.e., $m_{\star} =
1.56\,\msun$, $m_1 = m_2 = m_3 = 9\,\mJ$, $m_4 = 7\,\mJ$. The top panel presents the evolution of the astrocentric distances of the planets, while the bottom panel illustrates the evolution of the resonant angle of the double Laplace resonance, i.e., $\theta = \lambda_1 - 2\lambda_2 - \lambda_3 + 2\lambda_4$. {\em The middle and right-hand} columns: Final distribution of asteroids after $10$~Myr of evolution. The top panels present the position of asteroid in the orbital plane (dots), together with the positions of the planets (big dots). Gray rings show the temporal positions of the planets integrated over $100~$Myr after the migration stops. Black circle points the inner border of the outer debris disc of HR~8799 found in \citep{Booth2016}, i.e., $145\,\au$. Bottom panels present histograms of the asteroids' astrocentric distances. Vertical lines indicate the debris disc border (solid line) together with the uncertainties (dashed lines), i.e., $(145 \pm 12)\,\au$. The initial asteroids' eccentricities were chosen from the ranges of $[0,0.03]$ ({\em the middle} column) and $[0,0.3]$ ({\em the right} column).}
\label{fig:fig12}
\end{figure*}

All the constrains given above limit the initial orbits significantly. After a series of experiments we found an initial configuration and the migration parameters that fulfills the criteria. The evolution of the four-planet system into the resonant configuration is shown in {\em the left-hand} column of Fig.~\ref{fig:fig12}. The initial semi-major axes and other parameters are given in the caption of the figure. The outermost planet starts the migration at $90\,\au$, while the innermost one at $17.5\,\au$. The migration timescales between $9$ and~$28$~Myr (from the outermost to the innermost planets, respectively) together with the disc decay timescale of $1~$Myr make the planets to migrate inwards by a distance of $\sim 20\,\au$ for the outermost and $\sim 2\,\au$ for the innermost planets. The resonance (with low-amplitude libration of the resonant angle) is formed after $\sim 1~$Myr of the evolution, therefore the final system of four planets is long-term stable. 

The process of the resonant chain formation needs a comment. Entering into the resonant chain of four planets is a chaotic process, therefore even small changes in the initial configuration as well as differences between the numerical integrators used to solve the equations of motion may lead to different results, for instance the resonance could be not achieved, resulting in further self-disruption of the system. This is why the evolution shown here has to be treated as an example configuration. Initial configurations different to the one presented as well as different migration parameters may lead to qualitatively similar results.

Apart from the planets, there is a number of mass-less objects included in the model, whose motion is being perturbed by the planets. Their final distribution after $10$~Myr of the integration is shown in the middle and {\em the right-hand} columns of Fig.~\ref{fig:fig12}. We chose a uniform distribution of the particles semi-major axes, ranging from $15\,\au$ to $429\,\au$ \citep[the latter is the outer edge of the debris belt found by][]{Booth2016}. Initial eccentricities of the particles ranges from $0$ to $0.03$ ({\em the middle} column) and from $0$ to $0.3$ ({\rm the right-hand} column), all the angles are chosen randomly in the whole range of $(0^{\circ}, 360^{\circ})$. The final number of objects that remained in the system $\sim 10^5$ in each case. The inner edge of the asteroids belt places itself at $\sim 145\,\au$ for both ranges of the initial eccentricities. The border is more sharp for the simulations with initial $e<0.03$. Moreover, in this case there are spiral waves in the disc, while when initial $e<0.3$ the disc is axially symmetric. 

In both the cases, there is a small number of objects inside $145\,\au$, that results from non-rectangular radial distribution of the asteroids. We fine-tuned the initial orbits of the giant planets in such a way that at the distance of $145\,\au$ the radial density of the object equals approximately half of the radial density in the plateau region of the radial distribution. The border defined in this way can be easily controlled when the model with migration is applied, simply by shifting initial positions of the planets, as most of the asteroid are being removed from the system shortly after the beginning of the simulation, when the giant planets are still close to their initial positions.

%
\section{Discussion}
\label{sec:discussion}
Our results are derived under the major assumption that the orbital models of the \hr8799{} systems are resonant, and the strong, zero-th Laplace MMR chain is likely the primary factor maintaining the long-term stability. However, some recent results in the literature might apparently contradict this assumption, which possibly requires a comment.

\begin{figure}
\centerline{
\includegraphics[width=0.46\textwidth]{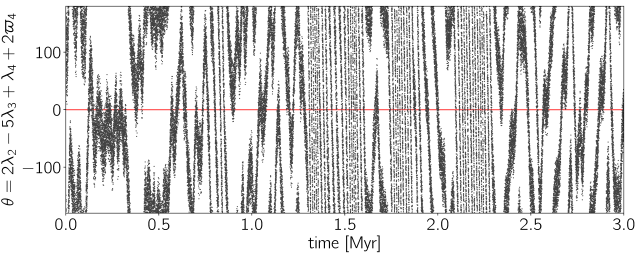}%
}
\caption{
Temporal evolution of a critical argument $\theta$ of the three-body 2d:-5e:1b~MMR of the system 94, simulation 4e in \citep[][their Appendix]{Gotberg2016}. A  sequence of alternating librations and rotations implies separatrix crossings and strongly chaotic evolution of the system.
} 
\label{fig:fig13}
\end{figure}

Recently, \cite{Gotberg2016} proposed long-term Lagrange-stable solutions found by tuning the initial orbital separations between the planets in terms of the Hill-radii spacing \citep{Chatterjee2008}. These long-living models are reported as non-resonant, strongly chaotic and prone to tiny changes of the initial conditions and a numerical scheme. 

We examined system 94 (simulation 4e) in Appendix of the source paper, given in the form of Cartesian, astrocentric coordinates. The frequency analysis of this solution reveals that it exhibits alternating rotations and librations of a critical argument of the three-body MMR 2d:-5c:1b (Fig.~\ref{fig:fig13}). This feature indicates {the} separatrix crossing. We also did 2-dim $\Ym$-scans in the $(a_\idm{\rm e},e_{\idm{\rm e}})$-plane, close to this solution (not shown), which reveal a dense net of other narrow multi-body MMRs in this region around $a_{\idm{\rm e}} \simeq 14.3$~au. The scans might be helpful to select other marginally stable solutions associated with narrow three-body and four-body MMRs of higher orders. 


The results of \cite{Gotberg2016} are consistent with our simulations in the sense that they indicate a marginal stability of the system. It may be long living only in very narrow stability regions in the phase space, associated with various multi-body MMRs. A resonance mechanism protecting the system from close encounters must be acting for the present planetary mass estimates, though the orbital evolution may be chaotic and marginally stable, as we also argue in \citep[][]{Gozdziewski2009,Gozdziewski2014A}, and in this work.

A determination of the Hill-radii separation through semi-major axes expressed  in astrocentric reference frame may be non-unique. For instance, the semi-major axis of \hr8799{}b in the model 94 (simulation 4e) in \citep{Gotberg2016} varies within $10$~au, depending on the orbital phase, although the orbit has been set initially almost circular. It may be explained by indirect perturbations in the astrocentric frame, see \citep[][]{Lee2001} for details; also this work, regarding identification of MMRs in the outer debris disk (Sects. \ref{sec:planetV} and \ref{sec:outer}). The osculating period of \hr8799{}b simultaneously varies roughly between 450 and 550~years that hinders the proper identification of the MMRs. The identification is much easier in the canonical reference frame. In that frame the $N$-body dynamics may be approximated to the first order in the masses through the Keplerian orbits perturbed by the mutual interactions between the planets \citep[e.g.,][]{Malhotra1993}.

A choice of the astrometric model is one more subtle, yet crucial factor for determining the long-term stable solutions. By reproducing the observations with Keplerian orbits, we dismiss information on the planet masses. Apparently, the mutual interactions should not be a significant factor {for resolving the orbital geometry} given that the observed arcs are at most $\sim 12\%$ of the full orbits. However, the masses are critical for determining the dynamical interactions in the system, hence skipping this prior seems to be inconsistent with the Bayesian inference. Even rigorously stable $N$-body models exhibit orbits which {\em do not close} after {\em just one} osculating period (Sect.~\ref{sec:best}). Unstable models marginally worse from mathematically best-fitting solutions are unconstrained and represent widely open arcs during only one {osculating period} (Fig.~\ref{fig:fig6}). These arguments indicate that the kinematic (geometric) models may be already biased, in spite of the short-arc measurements coverage.

%
%
\section{Summary}
\label{sec:conclusions}
In the first part of the paper, we present an updated astrometric model of the \hr8799{} planetary system. If the masses of detected planets are in a few Jupiter mass range, the system is strongly unstable in 1~Myr time-scale. It is still a very short interval of time, when compared to the star age, estimated between 30 and 160~Myr. Simulations of its dynamics conducted so far reveal that the long-term stable orbital architectures of the system  must be confined to small and particular regions in the phase space. We attempt to solve this paradox by assuming that the \hr8799{} system is involved in a 2-body MMR chain or multiple, three- or four-body MMR. Since the orbital parameters of a stable, resonant configuration are not independent and coupled, the number of free parameters of the astrometric model may be reduced to only five. This makes it possible to find stable solutions consistent with astrometric measurements and mass estimates derived from the planet formation and cooling theory.  

While the original version of this optimization method has been developed earlier, in this work we propose a structured and CPU-efficient algorithm that consists of two independent steps. At the first stage we build a database of stable configurations by simulating the planetary migration process. These synthetic, co-planar systems may agree only roughly with the spatial dimensions of the observed system. The second step performed with evolutionary algorithms is for fine-tuning these systems through the linear scaling, rotations of the orbits in space and propagating the bodies along their orbits  with the  $N$-body code to the proper, observed positions. We note that during the latter step, the elements may change self-consistently, in accord with the $N$-body evolution.

This approach lead us to finding the best-fitting model with $\Chi\simeq 1$, which corresponds to the MMR chain of a generalized, four-body Laplace resonance. This solution to a subset of self-consistent astrometric measurements in \citep{Konopacky2016} extrapolates to all observations to date and preserves the correct timing. In particular, the extrapolated orbit of planet \hr8799{}d passes closely to the first HST observation in 1998 \citep{Lafreniere2009}. {Solutions unconstrained by the migration are widely spread}, in spite of much smaller $\Chi \sim 0.74$, only marginally worse from mathematically best-fitting configuration
yielding $\Chi \sim 0.6$. It may be a strong argument supporting our resonant model. Also the mutual gravitational interactions between the planets are apparent since the best-fitting $N$-body orbits may collide in a time-scale of a few revolutions. That puts in concern kinematic models that do not account for the mutual interactions between the planets. 

In the second part, we used the stable best-fitting initial conditions for simulating the dynamical structure of debris disks in the system.  We developed CPU-efficient sampling of the initial conditions with the fast indicator MEGNO, dubbed as the $\Ym$-model. This dynamical model makes it also possible to locate additional, relatively massive objects in stable orbits, representing yet undetected planets in the system. The \hr8799{} system of four planets involved in the Laplace 8b:4c:2d:1e MMR is surprisingly robust for relatively strong perturbations introduced by such additional objects in 1-3~Jupiter mass range. Such unknown planets may be found interior or exterior to planets \hr8799{}e and  \hr8799{}b, respectively, and extending the MMR chain to five or more planets.

With the $\Ym$-model, we simulated the debris disks  composed of small asteroids. The structure of these disks is represented by temporal coordinates in the orbital plane, as well as by the canonical Keplerian elements in the $(a_{\rm p},e_{\rm p})$-plane. The simulations ended up with $\sim 10^6$ particles, and reveal regions which may be populated by asteroids or small planets in stable orbits. The total number of tested initial conditions was one-two orders of magnitude larger when counting unstable solutions. The $\Ym$-model was calibrated with the long-term, direct numerical integrations performed with the standard \mercury{} code.  The results for the four- and five-planet restricted problems derived with the direct numerical integrations for 34-68~Myr closely overlap with outcomes from the $\Ym$-model traced for much shorter intervals of $7$--$12$~Myr. 

The outer edge of the inner disk is shaped mostly by the inner planet~e. A border of stable motions may be roughly determined by the collision zone with this planet. Close to its orbit, stable orbits are permitted only when asteroids or small-mass objects are trapped in low-order MMRs. The overall image of this zone is similar for a range of masses, between $10^{-15}\mJ$ (small asteroids) and $1\mJ$ (Jovian planets). By allowing for the initial eccentricities of the test bodies up to the limit determined by collisional orbits, we detected stable resonances, like 3:2 and 1:1 MMRs missing in earlier papers \citep{Contro2016}. The presence of a significant proportion of asteroids in these resonances may contribute to highly non-symmetric edge of the inner disk.

We  also conducted CPU-intensive simulations of the outer disk, focusing on its inner edge and the inner part. Recently, it has been resolved with the ALMA observations in band 6 (1.34~mm) by \cite{Booth2016} and combined ALMA and VLA observations by \cite{Wilner2018}. They found the inner edge at 145$\pm 12$~au and 104$^{+8}_{-12}$~au, respectively. Followup, extensive simulations by \citep{Read2018} focus on the fifth, undetected planet carving the outer disk, to be consistent with the model of \cite{Booth2016}. Their best fitting model predicts a small 0.1$\mJ$ planet at $\simeq 138$~au. However, a different model in \citep{Wilner2018} may explain the ALMA and VLA observations with the currently observed four-body system. They also constrained a mass of planet \hr8799{}b to 5.8$^{+7.9}_{-3.1}$~au. 

We reconstructed the inner part of the disk composed of low-mass particles (asteroids) under different dynamical conditions, covering scenarios in these three papers, in the framework of the four- and five-planet restricted problem. 

The inner edge is mostly influenced by the outermost planet. The simulations ended up with $10^6$ bodies in stable orbits. The inner part of the outermost disk is spanned with various MMRs with this planet, including the 1:1 MMR and extended co-rotation zones. The width of these zones depends on the outermost planets' mass, and may extend for $\sim 100$~au. A border of stability region beyond the orbit of the outermost planet \hr8799{}b is determined by the collision zone with this planet. Low order resonances, like the 4:3, 3:2 and 2:1~MMR force eccentricity of the asteroids to moderate values. If these stable, moderate-eccentricity regions are populated, then the inner edge may exhibit very complex shape. It turns out to be more asymmetric for smaller masses of the outermost planet. We found the most complex edge for 0.1$\mJ$ at $\sim 138\mJ$, predicted in the best-fitting model to the ALMA observations in \citep{Read2018}. 

We note that in papers modeling the ALMA data, the edge is axi-symmetric with inclusion of the Lagrangian clumps appearing as non-important for the final results. However, our simulations indicate that the inner edge may be irregular, and particularly the 1:1~MMR zones overlapping with eccentric orbits in 3:2 and 4:3 MMRs may produce extended regions of emission. For instance, 0.33~$\mJ$ planet placed at $\sim 134$~au would be responsible for maintaining two huge Lagrangian clumps spanning approximately $120 \times 30$~au each. A smaller planet $\sim 0.1\mJ$ beyond $\sim 135$~au might {permit a complex structure of stable motions interior to its orbit and confined to low order MMRs with \hr8799b{}}.

Finally, we tentatively simulated the migration of the four observed planets in the presence of an extended asteroidal belt. We assume that the planets migrate to the final state of the 8b:4c:2d:1e Laplace resonance and influence the asteroids.  In this most complex astrophysical scenario, the disk edge may be moved to a ``desired'' location by appropriate migration rates and initial configuration of the planets. In such a case no additional planets would be necessary in order to explain the ALMA  observations. These simulations have a dynamic character, contrary to the static case considered in previous experiments. However, the static-type simulations made with the observed four-planet system are still complementary to the migration scenario. They reconstruct a detailed disk structure after the gaseous component has been dispersed, and the planet migration has slowed down or stopped. In particular, they reveal the dynamical structure of less populated regions between $\sim 90$ and $150$~au. 

We conclude that the \hr8799{} dynamical state cannot be yet fully resolved. While our resonant model may  explain the astrometric observations, a longer  time coverage is required to determine the real orbits without any doubt. As we confirm in this work, high-resolution images and observations of the debris disks may introduce additional, though indirect limits on the \hr8799{} system architecture. Various factors may be responsible for shaping the debris disks, like the presence of undetected planets, their masses and orbits. Therefore interpretation of the spectral energy distributions (SEDs) {and the spatial disk imaging} seems to be a complex and difficult problem, as the debris disk models may be non-unique. Our results and followup dynamical simulations, similar to those made in our paper, could be helpful to reduce this indeterminacy. 

\section*{Acknowledgements}

We thank Dan Fabrycky, the reviewer of our earlier paper \citep{Gozdziewski2014A}, for sharing with us the idea of optimizing the original \moa{} through the $N$-body dynamics scaling.
We thank the anonymous reviewer for comments which improved the work. This work has been supported by Polish National Science
Centre MAESTRO grant DEC-2012/06/A/ST9/00276. K.G. thanks the staff of the Pozna\'n Supercomputer and Network Centre (PCSS, Poland) for {the} generous long-term support and computing resources (grant No.~313).

\bibliographystyle{aasjournal}
\bibliography{ms}
\label{lastpage}
\end{document}